\documentclass[aps,prd,prabib,showpacs,nofootinbib]{revtex4}
\usepackage{graphicx} \usepackage{amsmath} \usepackage{amssymb}
\usepackage{amsfonts} \usepackage{bm}
\usepackage{array}
\usepackage{siunitx}
\usepackage[singlelinecheck=false]{caption}
\usepackage{ytableau}
\usepackage{multirow}

\newcommand*{\Comb}[2]{{}^{#1}C_{#2}}

\begin{document}

\newcommand{\be}{\begin{equation}} \newcommand{\ee}{\end{equation}}
\newcommand{\bea}{\begin{eqnarray}}\newcommand{\eea}{\end{eqnarray}}

\title{Non Self-conjugate   Strings, Singular Strings and  Rigged Configurations    in the  Heisenberg Model}

\author{Tetsuo Deguchi} \email{deguchi@phys.ocha.ac.jp} 

\author{Pulak Ranjan Giri} \email{pulakgiri@gmail.com}

\affiliation{ Department of Physics, Graduate School of Humanities and Sciences, Ochanomizu University, Ohtsuka 2-1-1, Bunkyo-ku, Tokyo, 112-8610, Japan}

\begin{abstract}
We observe a  different type of complex  solutions in the isotropic  spin-$1/2$ Heisenberg  chain starting from  $N=12$, where the central rapidity  of some of the  odd-length  strings  becomes complex making not all the strings  self-conjugate individually.  We show  that  there are at most $(N-2)/2$  singular    solutions  for    $M=4$, $M=5$ down-spins  and  at most $(N^2-6N+8)/8$ singular solutions for $M=6$, $M=7$ down-spins  in an even-length chain with $N \geq 2M$. Correspondence of the non  self-conjugate string  solutions   and the  singular string solutions   to  the rigged configurations  has also been shown.
\end{abstract}


\date{\today}

\maketitle

\section{Introduction}
Bethe's solution  of the isotropic spin-$1/2$ Heisenberg  model in one dimension,  by a method   known as the coordinate  Bethe ansatz \cite{bethe},  is one of the seminal works  in the field of integrable models.  For a detailed investigation  on the method and its variants and other related works see   references   \cite{gaud,faddeev,korepin,faddeev1,takab,nepo2,koma,mart,bei,suth}.  However, deriving  the  Bethe ansatz equations and the  form of the  eigenvalues and eigenvectors   is only one part of the story. The other important part is to extract the numerical values of the rapidities  from the set  of the Bethe ansatz equations.  Because of the  high degree of nonlinearity and multi-variate   nature   it is practically  impossible to analytically  solve the Bethe ansatz equations even for a  modest length of the spin chain. One, therefore needs to  seek  numerical solutions using methods, such as, iterations, the Newton-Raphson, homotopy continuations  etc. 
There have  been some efforts  to  find  the numerical solutions of the Bethe ansatz equations using different techniques,  and the eigenvalues have  been found which have excellent  match with  direct  diagonalization of the Hamiltonian.

Apart from the real solutions,  which are  much easier to find out,  there are  complex solutions of the Bethe ansatz equations,  which need extra efforts to calculate. Bethe himself investigated this problem for a  finite length  spin chain  and  found that  if there are  complex solutions then they come in complex conjugate pairs and arrange themselves  in  a string-like structure  \cite{bethe}.  These complex solutions, responsible for  the formation of  bound states,  lead to the so-called  {\it string hypothesis} \cite{taka}.   Importance  of  numerical computations of  all the Bethe solutions  are in one hand  to check  the completeness of the spectrum of the Hamiltonian and  on the other hand  the   knowledge of the numerical solutions  are  necessary  for the computation of   correlation functions,  form factors  \cite{maillet,maillet1} and other physical quantities of the model.

Although, the {\it string hypothesis}   gives satisfactory results  in the thermodynamic  limit and  counts  the total number of states correctly  in the general case,   it has many drawbacks and there have  been found  some exceptions  to it. For example,  if  the {\it string hypothesis} is  valid  in all respects then for a large length chain  the imaginary part of the rapidities of  $2$-string should converges  to $\pm 1/2$.  However it has been shown in \cite{vlad} that  some of the  $2$-string rapidities $\lambda_\pm$,  behave  as $\operatorname{Re}(\lambda_\pm) \sim N$, $\operatorname{Im}(\lambda_\pm)\sim \pm\sqrt{N}$ for large length $N$  spin chains.  Even, there are some $2$-strings which  for large $N$ and large Bethe quantum numbers deform back to two real rapidities \cite{essler,isler,ila,fuji}, which is  observed numerically. 
Despite these drawbacks, the {\it string hypothesis}  has been very helpful in numerical analysis in the iteration method  to obtain  good initial guess for the finite length chains.  Exploiting the {\it string hypothesis} and taking into account the deformations,  the complete string solutions for $N=8$ and $N=10$  length isotropic   spin-$1/2$  Heisenberg chains  are  obtained in \cite{hag}.    String solutions up to $N=14$ have been obtained  in \cite{nepo1}  and its supplement 
by making use of   the homotopy continuation method to show that there are too many solutions of the Bethe ansatz equations and only some of them, which obey the self-conjugacy condition,  are the physical solutions of the Heisenberg model. 
Here we remark that  we call  a solution of the Bethe ansatz equations {\it physical}  if it 
leads to an eigenvector of the Hamiltonian.

Usually, a set of  solutions to the Bethe ansatz equations consists  of strings of different lengths.  As mentioned above,  one key constraint to the solutions of the Bethe ansatz equations   is that they are self-conjugate  \cite{vlad2}.  While implementing  this constraint in the {\it string hypothesis} it is usually assumed that self-conjugacy is to be satisfied within  a  string \cite{hag}, a condition,  motivated by the observation of the  short length spin chains,  is too restrictive  to be valid for all solutions of  higher-length  spin chains.  We therefore  relax the imposition of the  self-conjugacy criterion  to the whole set of rapidities  not necessarily  within a string, making  the strings in a solution individually  not self-conjugate.  We show that  our  self-conjugacy criterion  allows us  to obtain  some  solutions which are not fitted within the standard deformed string picture, which is one of our motivations in this work.  For even-length chains,  up to  $N=10$  the string solutions,  although deformed, still do obey the  string structure and the restrictive self-conjugacy condition. First  breakdown of the string structure for the physical solutions  of even-length chains occurs  in  $N=12$,  as some of the  strings become non self-conjugate  and therefore need  the most relaxed self-conjugacy condition.  We discuss this feature of the string solutions here with an example of $N=12$.  Although numerical solutions for $N=12$ is obtained in \cite{nepo1} using the homotopy continuation method we here obtained the solutions  by the iteration method  using   Mathematica  and  exploiting the  {\it string hypothesis}. Here we remark that 
some different types of solutions of the Bethe ansatz equations in the anisotropic Heisenberg models are studied in \cite{woy,bab,fab1,fab2}.

Moreover, recently a lot of  works on the physical singular solutions have  been reported in the literature \cite{vlad1,sid,nepo,nepo1}.  The singular solutions as we know are an  essential component of the spectrum  and  need  a  proper regularization scheme to derive  the correct  physical states and the eigenvalues.  It is also possible to map these solutions  and even the regular solutions  to  a type  of combinatorial objects known as rigged configurations \cite{kirillov1,kirillov3,kirillov4}.  Based on the symmetry of the singular solutions  for even-length chains we  classify  the solutions  in different  categories,  which allow us to simplify  the Bethe ansatz  equations significantly up to $M=7$ down-spins.  Studying the  algebraic aspects  of  polynomials  we  estimate the number of singular solutions present at most for even-length spin chains  up to $M=7$ down-spins.   We also study the aspect of mapping of the Bethe states to rigged configurations for singular solutions as  well as   solutions with non self-conjugate strings.

Our study enables us to identify all the physical  singular solutions present in the $N=12$ case and map  them to rigged configurations. It also simplifies the Bethe ansatz equations for the singular solutions significantly  up to $M=7$ down-spins with even  $N$  so that the physical singular solutions  and their  total number  are obtained systematically.

We organize this paper in the following fashion:  In the next section,  we  briefly present the isotropic spin-$1/2$  chain and its solutions in terms of the algebraic Bethe ansatz method. In section III,  we discuss the non self-conjugate  strings    and explain  them  with the   example of  $N=12$.  In section IV,  we discuss the singular solutions, their  classification and give an  estimate  of the number singular solutions. In section V,  we discuss rigged  configurations and their  correspondence with $N=12$ case for non self-conjugate string solutions  and  singular string solutions   and finally we conclude.

\section{Algebraic Bethe Ansatz}
The Hamiltonian of a  spin-$1/2$ chain with length $N$  under the  periodic boundary conditions is given by  
\begin{eqnarray}\label{ham}
H= J\sum_{i=1}^{N}\left(S^x_iS^x_{i+1}+S^y_iS^y_{i+1}+S^z_iS^z_{i+1}- 1/4\right)\,,
\end{eqnarray}
where  $J$ is the coupling constant and $S_i^j (j=x,y,z)$  the  spin at position  $i$ and in $j$-direction.  In  the algebraic Bethe ansatz formulation one can construct a Bethe state  in the   case of $M$ down spin sector  as
\begin{eqnarray}\label{vec}
|\lambda_1, \lambda_2,\cdots,\lambda_M\rangle= \prod_{\alpha=1}^MB(\lambda_\alpha)|\Omega\rangle\,,
\end{eqnarray}
from the reference state $|\Omega\rangle$ with all up spins  by acting the  $B(\lambda_\alpha)$  matrix. To obtain the    $B(\lambda_\alpha)$  matrix we need the Lax operator 
\begin{eqnarray}
L_\gamma(\lambda)=\left( \begin{array}{cc}
\lambda-iS^z_\gamma & -iS^-_\gamma  \\
-iS^+_\gamma & \lambda+iS^z_\gamma  \end{array} \right)\,,
\end{eqnarray}
where $S^\pm_\gamma= S^x_\gamma \pm iS^y_\gamma$ are the Pauli  spin-$1/2$ matrixes. Each element of this matrix acts nontrivially  on the $\gamma$-th lattice site of the  Heisenberg model. One can  get  $B(\lambda_\alpha)$  from  the monodromy  matrix
\begin{eqnarray}
T(\lambda)=L_N(\lambda)L_{N-1}(\lambda) \cdots L_1(\lambda)=\left( \begin{array}{cc}
A(\lambda)& B(\lambda)  \\
C(\lambda) & D(\lambda) \end{array} \right)\,.
\end{eqnarray}
The Bethe  state (\ref{vec}) can also be written in a very useful form as  \cite{deguchi1}
\begin{eqnarray}\label{vec1}\nonumber
\prod_{\alpha=1}^MB(\lambda_\alpha)|\Omega\rangle= &&(-2i)^M \prod_{j < k}^{M}\frac{\lambda_j-\lambda_k +i}{\lambda_j-\lambda_k} \prod_{j =1}^{M}\frac{(\lambda_j-i/2)^N}{\lambda_j+i/2}\times \\
&&\sum_{1\leq x_1 < x_2......< x_M \leq N}^{N}\sum_{\mathcal{P}\in S_M}^{M!} \prod_{\mathcal{P}j <\mathcal{P}k}^{M} \left(\frac{\lambda_{\mathcal{P}j}-\lambda_{\mathcal{P}k}-i}{\lambda_{\mathcal{P}j}-\lambda_{\mathcal{P}k}+i}\right)^{H(j-k)} \prod_{j=1}^M\left(\frac{\lambda_{\mathcal{P}j}+i/2}{\lambda_{\mathcal{P}j}-i/2}\right)^{x_j}\prod_{j=1}^M S^-_{x_j}|\Omega \rangle\,,
\end{eqnarray}
where $\mathcal{P}$ are   elements of  the permutation  group $S_M $ of the $M$ numbers and $H(x)$ is the  Heaviside step function  $H(x) =1$ for $x >0$ and  $H(x) =0$ for $x \leq 0$.

When the rapidities  $\lambda_\alpha$  satisfy     the well known  Bethe ansatz equations 
\begin{eqnarray}\label{bethe}
\left(\frac{\lambda_\alpha- i/2}{\lambda_\alpha+ i/2}\right)^{N}= \prod_{\beta\neq \alpha}^{M}\frac{\lambda_\alpha-\lambda_\beta-i}{\lambda_\alpha-\lambda_\beta+i} ~~~  \alpha=1,2, \cdots, M\,,
\end{eqnarray}
then  the Bethe state given by  eq.  (\ref{vec})  ( and (\ref{vec1})) becomes   the highest-weight Bethe eigenstate. We also call the solutions of (\ref{bethe})  the Bethe roots.
The eigenvalue of the Hamiltonian (\ref{ham})  for the $M$ down-spin configuration is then  given by  
\begin{eqnarray}\label{gigen}
E= -J\frac{1}{2}\sum_{\alpha=1}^{M}\frac{1}{\left(\lambda_\alpha^2+ 1/4\right)}\,.
\end{eqnarray}
A convenient way to deal with  the Bethe ansatz  equations is to take the logarithm of  eq. (\ref{bethe}) 
  \begin{eqnarray}\label{logform}
2\arctan(2\lambda_\alpha)= J_\alpha\frac{2\pi}{N} + \frac{2}{N}\sum_{\beta\neq\alpha}^{M}\arctan(\lambda_\alpha-\lambda_\beta)~~~~\mbox{mod}~2\pi\,,
\end{eqnarray}
where the  Bethe quantum numbers, $\{J_\alpha, \alpha=1,2, \cdots, M\}$, take  integral or half integral values,  depending on whether  $N-M$ is odd or even, respectively.  However, since $J_\alpha$ are repetitive in  a given state, it is not useful for counting  the number of states  of the model in concern.   But, it is possible to  obtain   non-repetitive  quantum numbers  with the help of the  {\it string hypothesis}, which says that the rapidities for   $M$ down-spins   are arranged in a set of strings as 
\begin{eqnarray}\label{string}
\lambda_{\alpha a}^{j}= \lambda_{\alpha}^{j} +  \frac{i}{2}\left(j+1-2a\right) +  \Delta_{\alpha a}^{j} ~~~ a=1,2, \cdots, j, ~~\alpha=1,2, \cdots\,,
\end{eqnarray}
where the string center   $\lambda_\alpha^j$  is real, $j$ is the length of the string, $\alpha$  accounts for the number of  the $j$-strings  and  $\Delta_{\alpha a}^{j}$  are the string deviations.  In the limit  $\Delta_{\alpha a}^{j} \to 0$, the Bethe ansatz equations (\ref{bethe}) reduce to \cite{taka}
\begin{eqnarray}\label{betheta}\nonumber
\arctan\frac{2\lambda^j_\alpha}{j} &=& \pi \frac{I^j_\alpha}{N} + \frac{1}{N}\sum_{k=1}^{N_s}\sum_{\beta}^{M_k}\Theta_{jk}\left(\lambda_\alpha^j-\lambda_\beta^k\right)~~~~ \mbox{mod} ~\pi\,,\\
\Theta_{jk}(\lambda) &=& (1-\delta_{jk})\arctan\frac{2\lambda}{|j-k|} + 2\arctan\frac{2\lambda}{|j-k|+2} + \cdots + 2\arctan\frac{2\lambda}{j+k-2} + \arctan\frac{2\lambda}{j+k}\,,
\end{eqnarray}
where  $M_k$ is the number of  $k$-strings present in the $M$ down-spin state  and $N_s$ is the length of  the  largest string  such that $\sum_{k}^{N_s}kM_k=M$. The Takahashi quantum numbers $I^j_\alpha$, which are non-repetitive,   are given by
\begin{eqnarray}\label{takahashi} 
\mid I_{\alpha}^j\mid \leq \frac{1}{2} \left(N-1-\sum_{k=1}\left[2 ~\mbox{min}(j,k)-\delta_{j,k}\right]M_k\right)\,.
\end{eqnarray}

\begin{table}
\caption{Non self-conjugate Bethe roots for $N=12$, $M=5$. There are only $2$ solutions with  two  $1$-strings  and one $3$-string}
\begin{tabular}{l@{\hskip 0.2in}r@{\hskip 0.2in}r@{\hskip 0.2in}l@{\hskip 0.2in}l@{\hskip 0.2in}l@{\hskip 0.2in}}
\hline\hline
&$J$ & $I_n$ & \hspace{1.2in}$\lambda$& \hspace{0.6in} $E$ & Rigged Configuration\\
\hline
\multirow{5}*{1.}&\num{2}&$\num{3/2}_1$ &\num{0.18071431863183055} &\multirow{5}*{\num{-3.6006932562693255} }&
\multirow{5}*{\ytableausetup{centertableaux}
\begin{ytableau}
\none[2]& & & &\none[2] \\
\none[6]&  &\none[5] \\
\none[6]&  &\none[5] 
\end{ytableau}}\\
&\num{3}&$\num{5/2}_1$&\num{ 0.44476350644863927 - 0.01877019940237738 i} && \\
&\num{4} &\multirow{3}*{$\num{1}_3$}&  \num{0.49181421369589934 + 0.9614711323790809 i} & &\\
&\num{3}& &\num{0.44476350644863927 + 0.01877019940237738 i} & &\\
&\num{5}& &\num{0.49181421369589934 - 0.9614711323790809 i} & &\\
&&& & &\\
\multirow{5}*{2.}&\num{-2}&$\num{-3/2}_1$ & \num{-0.18071431863183055} &\multirow{5}*{\num{-3.6006932562693255} }&
\multirow{5}*{\ytableausetup{centertableaux}
\begin{ytableau}
\none[2]& & & &\none[0] \\
\none[6]&  &\none[1] \\
\none[6]&  &\none[1] 
\end{ytableau}}\\
&\num{-3}&$\num{-5/2}_1$&\num{ -0.44476350644863927 + 0.01877019940237738 i} && \\
&\num{-4}&\multirow{3}*{$\num{-1}_3$} & \num{-0.49181421369589934 - 0.9614711323790809 i} & &\\
&\num{-3} &&\num{-0.44476350644863927 - 0.01877019940237738 i} & &\\
&\num{-5}& &\num{-0.49181421369589934 + 0.9614711323790809 i} & &\\[1ex]
\hline
\end{tabular}
\end{table}

\begin{table}
\caption{Non self-conjugate Bethe roots for $N=12$, $M=6$.  There are three solutions with  one  $1$-string, one  $2$-string  and one $3$-string}
\begin{tabular}{l@{\hskip 0.2in}r@{\hskip 0.2in}r@{\hskip 0.2in}l@{\hskip 0.2in}l@{\hskip 0.2in}l@{\hskip 0.2in}}
\hline\hline
&$J$ & $I_n$ & \hspace{1.2in}$\lambda$& \hspace{0.6in} $E$ & Rigged Configuration\\
\hline
\multirow{6}*{1.}&\num{-1/2}&$\num{0}_1$&\num{ + 0.0185398999059037 i} & \multirow{6}*{ \num{-3.6497381892472}} &
\multirow{6}*{\ytableausetup{centertableaux}
\begin{ytableau}
\none[0]&  & & &\none[0] \\
\none[2]& & &\none[1] \\
\none[6]& &\none[3] 
\end{ytableau}}\\
&\num{5/2}&\multirow{2}*{$\num{0}_2$}&\num{+0.5i} & &\\
&\num{7/2}&&\num{-0.5i} & &\\
&\num{9/2}&\multirow{3}*{$\num{0}_3$} & \num{ +0.993775005875478 i} & & \\
&\num{1/2}&&\num{- 0.0185398999059037i} &&\\
&\num{-9/2} && \num{-0.993775005875478  i} & & \\
&&& &&\\
\multirow{6}*{2.}&\num{3/2}&$\num{2}_1$&\num{0.384905215843542 + 0.0190612670356019 i} & \multirow{6}*{\num{-2.46168458170981}} &
\multirow{6}*{\ytableausetup{centertableaux}
\begin{ytableau}
\none[0]&  & & &\none[0] \\
\none[2]& & &\none[0] \\
\none[6]& &\none[5] 
\end{ytableau}}\\
&\num{-7/2}&\multirow{2}*{$\num{-1}_2$}&\num{-0.752213256639834 + 0.507293831282871i} && \\
&\num{-5/2}&&\num{-0.752213256639834 - 0.507293831282871i} & &\\
&\num{9/2} &\multirow{3}*{$\num{0}_3$}& \num{0.367308040796292 + 0.991797190897116 i} & &\\
&\num{3/2}&&\num{0.384905215843542 - 0.0190612670356019 i} &&\\
&\num{11/2}& & \num{0.367308040796292 - 0.991797190897116 i} & &\\
&&&& &\\
\multirow{6}*{3.}&\num{-3/2}&$\num{-2}_1$&\num{-0.384905215843542 + 0.0190612670356019 i} & \multirow{6}*{\num{-2.46168458170981}} &
\multirow{6}*{\ytableausetup{centertableaux}
\begin{ytableau}
\none[0]&  & & &\none[0] \\
\none[2]& & &\none[2] \\
\none[6]& &\none[1] 
\end{ytableau}}\\
&\num{7/2}&\multirow{2}*{$\num{1}_2$}&\num{0.752213256639834 - 0.507293831282871i} && \\
&\num{5/2}&&\num{0.752213256639834 + 0.507293831282871i} & &\\
&\num{-9/2}&\multirow{3}*{$\num{0}_3$} & \num{-0.367308040796292 - 0.991797190897116 i} && \\
&\num{-3/2}&&\num{-0.384905215843542 - 0.0190612670356019 i}&&\\
&\num{-11/2} && \num{-0.367308040796292 + 0.991797190897116 i} && \\[1ex]
\hline
\end{tabular}
\end{table}

\begin{figure}[h!]
  \centering
    \includegraphics[width=0.45\textwidth]{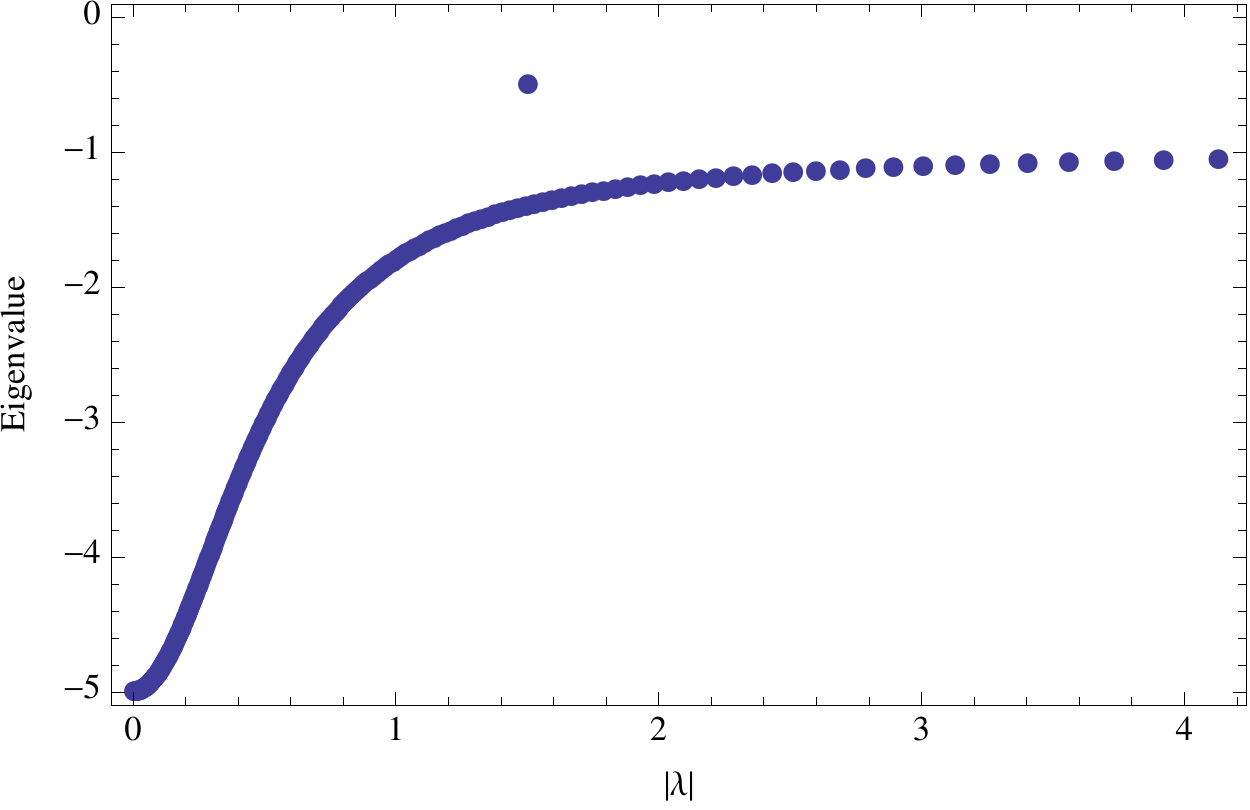}
  \caption{The energy  eigenvalues  for the singular solutions  for $N=500, M=4$ with absolute value  $|\lambda|$  of the form, $a_1$ or $ia_1$  of eqs. \ref{4sol1} and \ref{4sol2}, respectively. Up to $N=500$,  we obtained only one solution  of the form $ia_1$  for  every even-integer  $N\geq 8$, with $ia_1 \to 1.5i$ for large $N$ and the corresponding eigenvalue $E \to -0.5$ as can be seen from the isolated point in the figure}

\end{figure}

\begin{figure}[h!]
  \centering
    \includegraphics[width=0.45\textwidth]{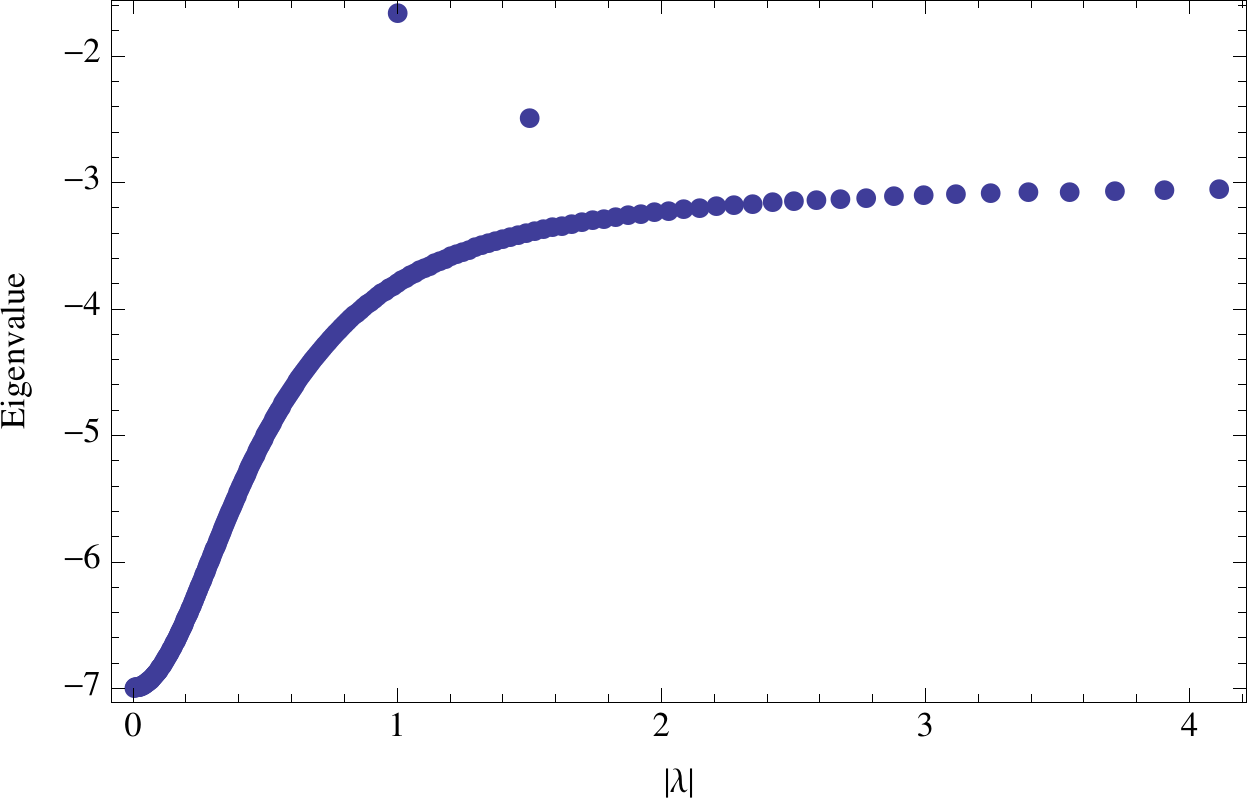}
    \caption{The energy eigenvalues  for the singular solutions  for $N=500, M=5$  with absolute value  $|\lambda|$  of the form, $a_1$ or $ia_1$  of eqs. \ref{5sol1} and \ref{5sol2}, respectively. Up to  $N=500$ we obtained only two solutions  of the form $ia_1$  for  every even $N \geq 10$, with $ia_1 \to 1.5i$ and $ia_1 \to 1i$, respectively  for large $N$ and the corresponding eigenvalues,  $E \to -2.5$  and $E \to -1.6666666666666667$,  respectively as can be seen from the two isolated points in the figure}
 
\end{figure}

\section{Non Self-conjugate  Strings}
In this section we present  the non-string type solutions, which start to occur from  the  $N=12$ case.   One of the important ingredients  for an effective iteration method is to  start the numerical  procedure  with a very good initial guess.  
In case of  the Bethe  equations (\ref{bethe}) it can be found by  solving the  Takahashi  equations (\ref{betheta}).   One then needs to find the deviations  $\Delta_{\alpha a}^{j}$  of the string  to obtain  the actual roots.  
In the  {\it string hypothesis} these deviations are  supposed to be  purely  imaginary  and decrease  exponentially with respect to $N$.  In finite-size spin chains, however,  there are deviations of the string center, which  leads  to the form  \cite{hag} 
\begin{eqnarray}\label{stringdev}
\Delta_{\alpha a}^{j}=  \epsilon_{\alpha a}^j + i\delta_{\alpha a}^j\,,
\end{eqnarray}
where $\epsilon_{\alpha a}^j $ and  $\delta_{\alpha a}^j$ are real.
Since the Bethe roots are self-conjugate there should be  a corresponding  restriction on the deviations.  One choice is to consider self-conjugacy  within a string  of length $j$, which  translates to  $\Delta_{\alpha a}^{j}= (\Delta_{\alpha  j+1-a}^{j})^*$. A consequence of this picture is that  the central rapidity of an odd-length string  is always  real and there is as such no relation between different strings in  an  eigenstate.  For  $N=8$ and $N=10$ we can recover all the string solutions from this  consideration of string picture and  it is therefore  sufficient to restrict  the self-conjugacy condition within a string.  However, as we increase the length $N$ of the spin chain   it is not  possible to recover  all the string solutions  since  the above  self-conjugacy condition is too restrictive if we   assume that deviations are  small.  Note that in the  {\it string hypothesis} it is crucial   to construct  the strings  in such  a way that the  deviations  (\ref{stringdev})  are small  enough so that the  $\Delta_{\alpha a}^{j} \to 0$ limit   leads to an approximate solution of the Bethe ansatz equations,  which satisfies  the Takahashi equations (\ref{betheta}) and gives  the Takahashi quantum numbers  (\ref{takahashi}). 
One  therefore needs to impose the self-conjugacy condition on the whole set of strings  \{$\lambda_{\alpha a}^{j}$\} = \{$(\lambda_{\alpha a}^{j})^*$\} as mentioned  in  introduction. Keeping the deviations small and  relaxing the self-conjugacy to the whole  set of strings, we obtain  some solutions, which are  not self-conjugate, we call  such  solutions  {\it non self-conjugate string} solutions.

To clarify the point with concrete examples let us consider a state of $M=5$  down-spins in a  $N=12$ spin chain, with two 1-strings  ($M_1=2$)  and one 3-string ($M_3=1$).   If we  consider the  self-conjugacy condition  only within each  string then the  Bethe roots can be parametrized as  
\begin{eqnarray}\label{5downspin}
 \{\lambda_1\},\{\lambda_2\},\{\lambda+\epsilon + (1+\delta)i,\lambda, \lambda+\epsilon - (1+\delta)i\}\,,
\end{eqnarray}
where $\lambda_1,\lambda_2,\lambda,\epsilon,\delta$ are real parameters which we have to evaluate numerically. We put curly brackets  to separate different strings.  All the physical solutions  for $N=12,M=5,M_1=2,M_3=1$  though fall in this category   except two solutions in which case the roots can be parametrized as
\begin{eqnarray}\label{5downspin1}
 \{\lambda_1\},\{\lambda +i\delta_1\},\{\lambda+\epsilon + (1+\delta_2)i,\lambda-i\delta_1, \lambda+\epsilon - (1+\delta_2)i\}\,,
\end{eqnarray}
where $\lambda_1,\lambda,\epsilon,\delta_1,\delta_2$ are real parameters.  In Table I   two  such  solutions of this kind  are shown where one of the  two  1-strings becomes complex conjugate to the central rapidity  of the three string and therefore  the  1-string and the 3-string  become non self-conjugate individually but remains self-conjugate when considered collectively.  In all the tables  $I_n$ represent  the Takahashi quantum numbers  for the  $n$-strings. 

Another example is present  in the $M=6$ down spin sector, with  one 1-string ($M_1=1$), one 2-string ($M_2=1$) and one 3-string ($M_3=1$). Again if we consider self-conjugacy for each string  then the  Bethe roots can be parametrized as
\begin{eqnarray}\label{6downspin}
 \{\lambda_2\},\{\lambda_1 +\frac{i}{2}(1+2\delta), \lambda_1 -\frac{i}{2}(1+2\delta)\},\{\lambda+\epsilon + (1+\delta_1)i,\lambda, \lambda+\epsilon - (1+\delta_1)i\}\,,
\end{eqnarray}
where $\lambda_1,\lambda_2,\lambda,\delta,\delta_1,\epsilon$ are real parameters. All the solutions fall  in  this category except the  three solutions which follow 
\begin{eqnarray}\label{6downspin1}
 \{\lambda +i\delta_1\},\{\lambda_1 +\frac{i}{2}(1+2\delta), \lambda_1 -\frac{i}{2}(1+2\delta)\},\{\lambda+\epsilon + (1+\delta_2)i,\lambda-i\delta_1, \lambda+\epsilon - (1+\delta_2)i\}\,,
\end{eqnarray}
where $\lambda,\lambda_1, \delta_2, \delta, \delta_1, \epsilon$ are real parameters. We can still use  (\ref{5downspin1}) and (\ref{6downspin1}) to get a relation between the Bethe quantum numbers  $J_i$ and the Takahashi quantum numbers  $I^\alpha_n$  by first simply   taking   $ \delta_1 \to 0$  limit  in the expressions of the two rapidities  $ \lambda + i\delta_1$  and $ \lambda - i\delta_1$.  We note that  in the   $ \delta_1 \to 0$  limit  the two rapidities do not become equal to each other,  rather they  go to two  distinct  rapidities,   
 $\lim_{\delta_1\to 0} \lambda  + i\delta_1 \to  \lambda_2$ and   $\lim_{\delta_1\to 0} \lambda  - i\delta_1 \to  \lambda$.  
Then  (\ref{5downspin1}) and (\ref{6downspin1})  reduces to the standard string roots of  (\ref{5downspin}) and (\ref{6downspin}) respectively, which allows us to obtain  the  Takahashi  quantum numbers in terms of the Bethe quantum numbers. 
In Table II the  three  solutions  of the form (\ref{6downspin1})  have been shown.

\begin{table}
\caption{Singular Bethe roots for $N=12$, $M=2$. There is one solution with one $2$-string}
\begin{tabular}{l@{\hskip 0.2in}r@{\hskip 0.2in}l@{\hskip 0.2in}l@{\hskip 0.2in}l@{\hskip 0.2in}l@{\hskip 0.2in}}
\hline\hline
&$J$& $I_n$ & $\lambda$& \hspace{0.1in} $E$ & Rigged Configuration\\
\hline
\multirow{2}*{1.}&\num{5/2}&\multirow{2}*{$\num{0}_2$} & \num{0.5i} & \multirow{2}*{\num{ -1.}} &
\multirow{2}*{\ytableausetup{centertableaux}
\begin{ytableau}
\none[8]&  & &\none[4] 
\end{ytableau}}\\
&\num{7/2}&&\num{ -0.5i} & &\\[1ex]
\hline
\end{tabular}
\end{table}

\begin{table}
\caption{Singular Bethe roots for $N=12$, $M=3$. There is one solution with  one $1$-string  and one  $2$-string}
\begin{tabular}{l@{\hskip 0.2in}l@{\hskip 0.2in}l@{\hskip 0.2in}l@{\hskip 0.2in}l@{\hskip 0.2in}l@{\hskip 0.2in}}
\hline\hline
&$J$ & $I_n$& $\lambda$&\hspace{0.1in} $E$ & Rigged Configuration\\
\hline
\multirow{3}*{1.}&\num{0} & $\num{0}_1$& \num{0.} & \multirow{3}*{\num{-3.0}}&
\multirow{3}*{\ytableausetup{centertableaux}
\begin{ytableau}
\none[6]&  & &\none[3] \\
\none[8] & &\none[4] 
\end{ytableau}}\\
&\num{3} &\multirow{2}*{$\num{0}_2$}&\num{ 0.5i} & &\\
&\num{3}& &\num{-0.5i}& &\\[1ex]
\hline
\end{tabular}
\end{table}

\begin{table}
\caption{Singular Bethe roots for $N=12$, $M=4$. There are four  solutions with  two $1$-strings   and one $2$-string}
\begin{tabular}{l@{\hskip 0.2in}r@{\hskip 0.2in}r@{\hskip 0.2in}l@{\hskip 0.2in}l@{\hskip 0.2in}l@{\hskip 0.2in}}
\hline\hline
&$J$ & $I_n$ & \hspace{0.6in}$\lambda$&\hspace{0.6in} $E$& Rigged Configuration \\
\hline
\multirow{4}*{1.}&\num{-7/2} &$\num{-7/2}_1$& \num{-1.1657640741782098} &\multirow{4}*{\num{-1.6215017698290826 } }&
\multirow{4}*{\ytableausetup{centertableaux}
\begin{ytableau}
\none[4]&  & &\none[2] \\
\none[6] & &\none[0] \\
\none[6] & &\none[6]
\end{ytableau}}\\
&\num{7/2}&$\num{7/2}_1$ & \num{1.1657640741782098} && \\
&\num{5/2}&\multirow{2}*{$\num{0}_2$}&\num{ 0.5i} & &\\
&\num{7/2}&&\num{ -0.5i} & &\\
&&&&\\
\multirow{4}*{2.}&\num{-5/2} &$\num{-5/2}_1$& \num{-0.535523144483441} & \multirow{4}*{\num{-2.8629431312188682}} &
\multirow{4}*{\ytableausetup{centertableaux}
\begin{ytableau}
\none[4]&  & &\none[2] \\
\none[6] & &\none[1] \\
\none[6] & &\none[5]
\end{ytableau}}\\
&\num{5/2} &$\num{5/2}_1$& \num{0.535523144483441} & &\\
&\num{5/2}&\multirow{2}*{$\num{0}_2$}&\num{ 0.5i} & & \\
&\num{7/2}&&\num{ -0.5i} & & \\
&&&&\\
\multirow{4}*{3.}&\num{-3/2} &$\num{-3/2}_1$& \num{-0.2659137285956084} &\multirow{4}*{ \num{-4.118080676373112}} &
\multirow{4}*{\ytableausetup{centertableaux}
\begin{ytableau}
\none[4]&  & &\none[2] \\
\none[6] & &\none[2] \\
\none[6] & &\none[4]
\end{ytableau}}\\
&\num{3/2} &$\num{3/2}_1$& \num{0.2659137285956084} & &\\
&\num{5/2}&\multirow{2}*{$\num{0}_2$}&\num{ 0.5i} & &\\
&\num{7/2}&&\num{ -0.5i} & &\\
&&&&\\
\multirow{4}*{4.}&\num{-1/2} &$\num{-1/2}_1$& \num{-0.08199356634008453} & \multirow{4}*{\num{-4.895249800546962}}&
\multirow{4}*{\ytableausetup{centertableaux}
\begin{ytableau}
\none[4]&  & &\none[2] \\
\none[6] & &\none[3] \\
\none[6] & &\none[3]
\end{ytableau}}\\
&\num{1/2} &$\num{1/2}_1$& \num{0.08199356634008453} & &\\
&\num{5/2}&\multirow{2}*{$\num{0}_2$}&\num{ 0.5i} & &\\
&\num{7/2}&&\num{ -0.5i} & &\\[1ex]
\hline
\end{tabular}
\end{table}

\begin{table}
\caption{Singular Bethe roots for $N=12$, $M=4$. There is one solution with  one $4$-string}
\begin{tabular}{l@{\hskip 0.2in}r@{\hskip 0.2in}l@{\hskip 0.2in}l@{\hskip 0.2in}l@{\hskip 0.2in}l@{\hskip 0.2in}}
\hline\hline
&$J$& $I_n$ & \hspace{0.6in}$\lambda$& \hspace{0.6in} $E$& Rigged Configuration \\
\hline
\multirow{4}*{1.}&\num{9/2} &\multirow{4}*{$\num{0}_4$}& \num{1.502976465754898i} &\multirow{4}*{ \num{-0.5022246220319766 }} &
\multirow{4}*{\ytableausetup{centertableaux}
\begin{ytableau}
\none[4]&  && & &\none[2] 
\end{ytableau}}\\
&\num{5/2}&&\num{ 0.5i} & &\\
&\num{7/2}&&\num{ -0.5i} & &\\
&\num{-9/2} & &\num{-1.502976465754898i} &&\\[1ex]
\hline
\end{tabular}
\end{table}

\begin{table}
\caption{Singular Bethe roots for $N=12$, $M=5$. There are three solutions with  three $1$-strings   and one $2$-string}
\begin{tabular}{l@{\hskip 0.2in}r@{\hskip 0.2in}r@{\hskip 0.2in}l@{\hskip 0.2in}l@{\hskip 0.2in}l@{\hskip 0.2in}}
\hline\hline
&$J$ & $I_n$&\hspace{0.6in}$\lambda$& \hspace{0.6in} $E$ & Rigged Configuration\\
\hline
\multirow{5}*{1.}&\num{0}&$\num{0}_1$  & \num{0.} &\multirow{5}*{ \num{-3.9168598958278036}} &
\multirow{5}*{\ytableausetup{centertableaux}
\begin{ytableau}
\none[2]&  & &\none[1] \\
\none[4] & &\none[0] \\
\none[4] & &\none[2]\\
\none[4] & &\none[4]
\end{ytableau}}\\
&\num{-3}&$\num{-3}_1$  & \num{-0.9168855875745959} & & \\
&\num{3}&$\num{3}_1$  & \num{0.9168855875745959} &  &\\
&\num{3}&\multirow{2}*{$\num{0}_2$} &\num{ 0.5i} & &\\
&\num{3}&&\num{ -0.5i} & &\\
&&&&&\\
\multirow{5}*{2.}&\num{0}&$\num{0}_1$  & \num{0.} & \multirow{5}*{\num{-5.3599633111443925}} &
\multirow{5}*{\ytableausetup{centertableaux}
\begin{ytableau}
\none[2]&  & &\none[1] \\
\none[4] & &\none[1] \\
\none[4] & &\none[2]\\
\none[4] & &\none[3]
\end{ytableau}}\\
&\num{-2}&$\num{-2}_1$  & \num{-0.4168157878523103} &  &\\
&\num{2} &$\num{2}_1$ & \num{0.4168157878523103} &  &\\
&\num{3}&\multirow{2}*{$\num{0}_2$} &\num{ 0.5i} & &\\
&\num{3}&&\num{ -0.5i} & &\\
&&&&&\\
\multirow{5}*{3.}&\num{0}&$\num{0}_1$  & \num{0.} &\multirow{5}*{\num{-6.545681807497121}} &
\multirow{5}*{\ytableausetup{centertableaux}
\begin{ytableau}
\none[2]&  & &\none[1] \\
\none[4] & &\none[2] \\
\none[4] & &\none[2]\\
\none[4] & &\none[2]
\end{ytableau}}\\
&\num{-1}&$\num{-1}_1$  & \num{-0.1789782217190064} & & \\
&\num{1}&$\num{1}_1$  & \num{0.1789782217190064} &  &\\
&\num{3}&\multirow{2}*{$\num{0}_2$} &\num{ 0.5i} & &\\
&\num{3}&&\num{ -0.5i} & &\\[3ex]
\hline
\end{tabular}
\end{table}

\begin{table}
\caption{Singular Bethe roots for  $N=12$, $M=5$. There is one solution with one $1$-string  and one $4$-string}
\begin{tabular}{l@{\hskip 0.2in}r@{\hskip 0.2in}l@{\hskip 0.2in}l@{\hskip 0.2in}l@{\hskip 0.2in}l@{\hskip 0.2in}}
\hline\hline
&$J$ & $I_n$ & \hspace{0.6in}$\lambda$& \hspace{0.6in} $E$ & Rigged Configuration \\
\hline
\multirow{5}*{1.}&\num{0} &$\num{0}_1$& \num{0.} & \multirow{5}*{\num{-2.511429026296249} }&
\multirow{5}*{\ytableausetup{centertableaux}
\begin{ytableau}
\none[2]&  && & &\none[1] \\
\none[8]& &\none[4] 
\end{ytableau}}\\
&\num{4}&\multirow{4}*{$\num{0}_4$} & \num{1.5155149393260654i} & &\\
&\num{3}&&\num{ 0.5i} & &\\
&\num{3}&&\num{ -0.5i} & &\\
&\num{-4} & & \num{-1.5155149393260654i} & &\\[1ex]
\hline
\end{tabular}
\end{table}

\begin{table}
\caption{Singular Bethe roots for $N=12$, $M=5$. There is one solution with  one $2$-string  and one $3$-string}
\begin{tabular}{l@{\hskip 0.2in}r@{\hskip 0.2in}l@{\hskip 0.2in}l@{\hskip 0.2in}l@{\hskip 0.2in}l@{\hskip 0.2in}}
\hline\hline
& $J$ & $I_n$ & \hspace{0.6in}$\lambda$& \hspace{0.6in} $E$ & Rigged Configuration \\
\hline
\multirow{5}*{1.}&\num{3}&\multirow{2}*{$\num{0}_2$} &\num{ 0.5i} & \multirow{5}*{\num{-1.6660659592344356}} &
\multirow{5}*{\ytableausetup{centertableaux}
\begin{ytableau}
\none[2]&  & & &\none[1] \\
\none[4]& & &\none[2] 
\end{ytableau}}\\
&\num{3}&&\num{ -0.5i} & &\\
&\num{0}&\multirow{3}*{$\num{0}_3$} & \num{0.} & &\\
&\num{5}& & \num{0.9998311128556481i} &  &\\
&\num{-5} & & \num{-0.9998311128556481i } & &\\[1ex]
\hline
\end{tabular}
\end{table}

\begin{table}
\caption{Singular Bethe roots for  $N=12$, $M=6$. There are three solutions with four $1$-strings   and one $2$-string}
\begin{tabular}{l@{\hskip 0.2in}r@{\hskip 0.2in}r@{\hskip 0.2in}l@{\hskip 0.2in}l@{\hskip 0.2in}l@{\hskip 0.2in}}
\hline\hline
&$J$ & $I_n$ &\hspace{0.6in} $\lambda$& \hspace{0.6in}$E$ & Rigged Configuration\\
\hline
\multirow{6}*{1.}&\num{-5/2}&$\num{-5/2}_1$&\num{-0.6905065538178187} & \multirow{6}*{\num{-5.352050317651034}}&
\multirow{6}*{\ytableausetup{centertableaux}
\begin{ytableau}
\none[0]&  & &\none[0] \\
\none[2] & &\none[0] \\
\none[2] & &\none[0]\\
\none[2] & &\none[2]\\
\none[2] & &\none[2]
\end{ytableau}}\\
&\num{-3/2}&$\num{-3/2}_1$&\num{-0.29326445955096875} & &\\
&\num{3/2}&$\num{3/2}_1$&\num{0.29326445955096875} & &\\
&\num{5/2}&$\num{5/2}_1$&\num{0.6905065538178187} &  &\\
&\num{5/2}&\multirow{2}*{$\num{0}_2$}&\num{ 0.5i} & &\\
&\num{7/2} && \num{-0.5i} & &\\
&&&&&\\
\multirow{6}*{2.}&\num{-5/2}&$\num{-5/2}_1$&\num{-0.7003461585874278} & \multirow{6}*{\num{-6.229463841147066}}&
\multirow{6}*{\ytableausetup{centertableaux}
\begin{ytableau}
\none[0]&  & &\none[0] \\
\none[2] & &\none[0] \\
\none[2] & &\none[1]\\
\none[2] & &\none[1]\\
\none[2] & &\none[2]
\end{ytableau}}\\
&\num{-1/2}&$\num{-1/2}_1$&\num{-0.08830964234306034} & &\\
&\num{1/2}&$\num{1/2}_1$&\num{0.08830964234306034} & &\\
&\num{5/2}&$\num{5/2}_1$&\num{0.7003461585874278} & &\\
&\num{5/2}&\multirow{2}*{$\num{0}_2$}&\num{ 0.5i} & &\\
&\num{7/2} && \num{-0.5i} & &\\
&&&&&\\
\multirow{6}*{3.}&\num{-3/2}&$\num{-3/2}_1$&\num{-0.30694160344558236} &\multirow{6}*{\num{-7.77738933370129}}&
\multirow{6}*{\ytableausetup{centertableaux}
\begin{ytableau}
\none[0]&  & &\none[0] \\
\none[2] & &\none[1] \\
\none[2] & &\none[1]\\
\none[2] & &\none[1]\\
\none[2] & &\none[1]
\end{ytableau}}\\
&\num{-1/2}&$\num{-1/2}_1$&\num{-0.09083103807287891} & &\\
&\num{1/2}&$\num{1/2}_1$&\num{0.09083103807287891} & &\\
&\num{3/2}&$\num{3/2}_1$&\num{0.30694160344558236} & &\\
&\num{5/2}&\multirow{2}*{$\num{0}_2$}&\num{ 0.5i} & & \\
&\num{7/2} && \num{-0.5i} & & \\ [3ex]
\hline
\end{tabular}
\end{table}

\begin{table}
\caption{Singular Bethe roots for $N=12, M=6$.  There are four  solutions with two $1$-strings   and one $4$-string}
\begin{tabular}{l@{\hskip 0.2in}r@{\hskip 0.2in}r@{\hskip 0.2in}l@{\hskip 0.2in}l@{\hskip 0.2in}l@{\hskip 0.2in}}
\hline\hline
&$J$ & $I_n$ & \hspace{0.6in}$\lambda$& \hspace{0.6in}$E$ & Rigged Configuration\\
\hline
\multirow{6}*{1.}&\num{-5/2}&$\num{-7/2}_1$&\num{-0.9468899269636574} & \multirow{6}*{\num{-1.3869684651709449}}&
\multirow{6}*{\ytableausetup{centertableaux}
\begin{ytableau}
\none[0]&  & & & &\none[0] \\
\none[6]& &\none[0] \\
\none[6]& &\none[6] 
\end{ytableau}}\\
&\num{5/2}&$\num{7/2}_1$&\num{0.9468899269636574} & &\\
&\num{9/2}&\multirow{4}*{$0_4$}&\num{1.5202343808518264i} & &\\
&\num{5/2}&&\num{ 0.5i} & &\\
&\num{7/2}&&\num{ -0.5i} & &\\
&\num{-9/2}&&\num{-1.5202343808518264i} & &\\
&&&&&\\
\multirow{6}*{2.}&\num{-3/2}&$\num{-5/2}_1$&\num{-0.4781191178586541} & \multirow{6}*{\num{-2.623262633690177}}&
\multirow{6}*{\ytableausetup{centertableaux}
\begin{ytableau}
\none[0]&  & & & &\none[0] \\
\none[6]& &\none[1] \\
\none[6]& &\none[5] 
\end{ytableau}}\\
&\num{3/2}&$\num{5/2}_1$&\num{0.4781191178586541} & &\\
&\num{9/2}&\multirow{4}*{$0_4$}&\num{1.547617992727387 i} & &\\
&\num{5/2}&&\num{ 0.5i} & &\\
&\num{7/2}&&\num{ -0.5i} & &\\
&\num{-9/2}&&\num{-1.547617992727387 i} & &\\
&&&&&\\
\multirow{6}*{3.}&\num{-1/2}&$\num{-3/2}_1$&\num{-0.24554120504274426} & \multirow{6}*{\num{-3.773151385638714}}&
\multirow{6}*{\ytableausetup{centertableaux}
\begin{ytableau}
\none[0]&  & & & &\none[0] \\
\none[6]& &\none[2] \\
\none[6]& &\none[4] 
\end{ytableau}}\\
&\num{1/2}&$\num{3/2}_1$&\num{0.24554120504274426} & &\\
&\num{9/2}&\multirow{4}*{$0_4$}&\num{1.5729034567782043 i} & &\\
&\num{5/2}&&\num{ 0.5i} & &\\
&\num{7/2}&&\num{ -0.5i} & &\\
&\num{-9/2}&&\num{-1.5729034567782043i} & &\\
&&&&&\\
\multirow{6}*{4.}&\num{1/2}&$\num{-1/2}_1$&\num{-0.07666936301920817} & \multirow{6}*{\num{-4.46706634996716}}&
\multirow{6}*{\ytableausetup{centertableaux}
\begin{ytableau}
\none[0]&  & & & &\none[0] \\
\none[6]& &\none[3] \\
\none[6]& &\none[3] 
\end{ytableau}}\\
&\num{-1/2}&$\num{1/2}_1$&\num{0.07666936301920817} &&\\
&\num{9/2}&\multirow{4}*{$0_4$}&\num{1.5866164135168153 i} &&\\
&\num{5/2}&&\num{ 0.5i} &&\\
&\num{7/2}&&\num{ -0.5i} & &\\
&\num{-9/2}&&\num{-1.5866164135168153 i} &&\\[3ex]
\hline
\end{tabular}
\end{table}

\begin{table}
\caption{Singular Bethe roots for $N=12, M=6$. There is one solution with one $1$-string, one $2$-string  and  one $3$-string}
\begin{tabular}{l@{\hskip 0.2in}r@{\hskip 0.2in}l@{\hskip 0.2in}l@{\hskip 0.2in}l@{\hskip 0.2in}l@{\hskip 0.2in}}
\hline\hline
&$J$ & $I_n$ & \hspace{0.6in}$\lambda$& \hspace{0.6in}$E$ & Rigged Configuration\\
\hline
\multirow{6}*{1.}&\num{-1/2}&$\num{0}_1$&\num{0.018539899905903653i} & \multirow{6}*{\num{-3.649738189247195}}&
\multirow{6}*{\ytableausetup{centertableaux}
\begin{ytableau}
\none[0]&  & & &\none[0] \\
\none[2]& & &\none[1] \\
\none[6]& &\none[3] 
\end{ytableau}}\\
&\num{5/2}&\multirow{2}*{$0_2$}&\num{ 0.5i} & &\\
&\num{7/2}&&\num{ -0.5i} & &\\
&\num{9/2}&\multirow{3}*{$0_3$}&\num{0.9937750058754778 i} &&\\
&\num{1/2}&&\num{-0.018539899905903653i} &&\\
&\num{-9/2}&&\num{-0.9937750058754778 i} &&\\[1ex]
\hline
\end{tabular}
\end{table}

\begin{table}
\caption{Singular Bethe roots for $N=12, M=6$. There is one solution with three $2$-strings}
\begin{tabular}{l@{\hskip 0.2in}r@{\hskip 0.2in}r@{\hskip 0.2in}l@{\hskip 0.2in}l@{\hskip 0.2in}l@{\hskip 0.2in}}
\hline\hline
&$J$ & $I_n$ & \hspace{1.2in}$\lambda$&\hspace{0.6in} $E$ & Rigged Configuration \\
\hline
\multirow{6}*{1.}&\num{-7/2}&\multirow{2}*{$\num{-1}_2$}&\num{-0.6620239184153354+ 0.5045174233098804i} & \multirow{6}*{\num{-2.367482833109191}}&
\multirow{6}*{\ytableausetup{centertableaux}
\begin{ytableau}
\none[0]& & &\none[0] \\
\none[0]& & &\none[0] \\
\none[0]& & &\none[0] 
\end{ytableau}}\\
&\num{-5/2}&&\num{-0.6620239184153354- 0.5045174233098804i} &&\\
&\num{5/2}&\multirow{2}*{$0_2$}&\num{ 0.5i} &&\\
&\num{7/2}&&\num{ -0.5i} & &\\
&\num{5/2}&\multirow{2}*{$\num{1}_2$}&\num{0.6620239184153354+ 0.5045174233098804i} &&\\
&\num{7/2}&&\num{0.6620239184153354- 0.5045174233098804i} &&\\[1ex]
\hline
\end{tabular}
\end{table}

\begin{table}
\caption{Singular Bethe roots for $N=12$, $M=6$. There is one solution with  one $6$-string}
\begin{tabular}{l@{\hskip 0.2in}r@{\hskip 0.2in}l@{\hskip 0.2in}l@{\hskip 0.2in}l@{\hskip 0.2in}l@{\hskip 0.2in}}
\hline\hline
&$J$ & $I_n$ & \hspace{0.6in}$\lambda$& \hspace{0.6in}$E$ & \hspace{0.2in}Rigged Configuration\\
\hline
\multirow{6}*{1.}&\num{7/2}&\multirow{6}*{$0_6$}&\num{2.849226471551315i} & \multirow{6}*{\num{-0.3734266506772359}}&
\multirow{6}*{\ytableausetup{centertableaux}
\begin{ytableau}
\none[0]& & & & & & &\none[0] 
\end{ytableau}}\\
&\num{9/2}&&\num{1.500696932968625i} & &\\
&\num{5/2}&&\num{ 0.5i} & &\\
&\num{7/2}&&\num{ -0.5i} & &\\
&\num{-9/2}&&\num{-1.500696932968625i} & &\\
&\num{-7/2}&&\num{-2.849226471551315i} & &\\[1ex]
\hline
\end{tabular}
\end{table}

\section{Singular Strings}
Singular string solutions of the Bethe equations are  special in the sense  that  the energy  eigenvalue  diverges  and the Bethe state  vanishes  without regularization   and  therefore we need to have a suitable regularization scheme \cite{bei,vlad1,nepo,sid,kirillov2} 
to make everything  finite.  It is also an essential  part of the spectrum because without the singular solutions the  Hilbert space of the Hamiltonian is not complete.  Recently there has  been much  interest in singular solutions  and it is also possible to map all the singular solutions to  rigged configurations.   Solutions  of the form
\begin{eqnarray}\label{sgnrap}
\Big\{\lambda_1=\frac{i}{2},\lambda_2= -\frac{i}{2}, \lambda_3,\lambda_4, \cdots, \lambda_M \Big\}\,,
\end{eqnarray} 
are called singular because  two of the roots $\lambda_1, \lambda_2$ make the state  and the corresponding eigenvalue ill-defined.  It has been numerically observed that  the roots of a singular state for even $N$  are distributed symmetrically,  we assume  that  for the  physical singular solutions the following condition is satisfied
\begin{eqnarray}\label{singcond}
\sum_{\alpha=1}^M\lambda_\alpha= 0\,.
\end{eqnarray}
The condition (\ref{singcond}) is satisfied for any symmetrically distributed roots, which may or may not be singular.  We remark that  the condition  (\ref{singcond})  has been conjectured in   \cite{kirillov} for even length spin chain.
Another point  is  that the  singular solutions in \cite{nepo} for even $N$, which  do not satisfy the condition (\ref{singcond}),   are not  physical.   
In  singular 2-string    $\{\pm \frac{i}{2}\}$  case  the Bethe eigenstate (\ref{vec1})   takes a  simple form \cite{essler,nepo1}
\begin{eqnarray}\label{sgnvec1}
|\Psi\rangle_2 \equiv  \sum_{j=1}^N (-1)^j S^-_j S^-_{j+1}|\Omega\rangle\,.
\end{eqnarray}
Based on  (\ref{singcond}) and the self-conjugacy condition we can classify the different  singular states for a fixed  number of down   spins.  For $M=2$, the only  singular solution is  of the form 
\begin{eqnarray}\label{2sol1}
\Big\{\frac{i}{2},-\frac{i}{2}\Big\}\,,
\end{eqnarray}
which is given  for $N=12$ in table III. For $M=3$ the only singular solution possible is  
\begin{eqnarray}\label{3sol1}
\Big\{\frac{i}{2},-\frac{i}{2}, 0\Big\}\,,
\end{eqnarray}
which is given  in table IV.  Note that for $M=2$ and $M=3$, there is only one singular state for any even $N\geq 4$. For $M=4$, there are two different classes  of singular solutions 
 \begin{eqnarray}\label{4sol1}
&&\Big\{\frac{i}{2},-\frac{i}{2}, a_1, -a_1\Big\} ~~~~~~ \mbox{for}~~~ a_1 \in \mathbb{R}\,,\\ \label{4sol2}
&&\Big\{\frac{i}{2},-\frac{i}{2},  ia_1, -ia_1\Big\} ~~~~ \mbox{for}~~~ a_1\in \mathbb{R}\,.
\end{eqnarray}
\begin{figure}[h!]
  \centering
    \includegraphics[width=0.45\textwidth]{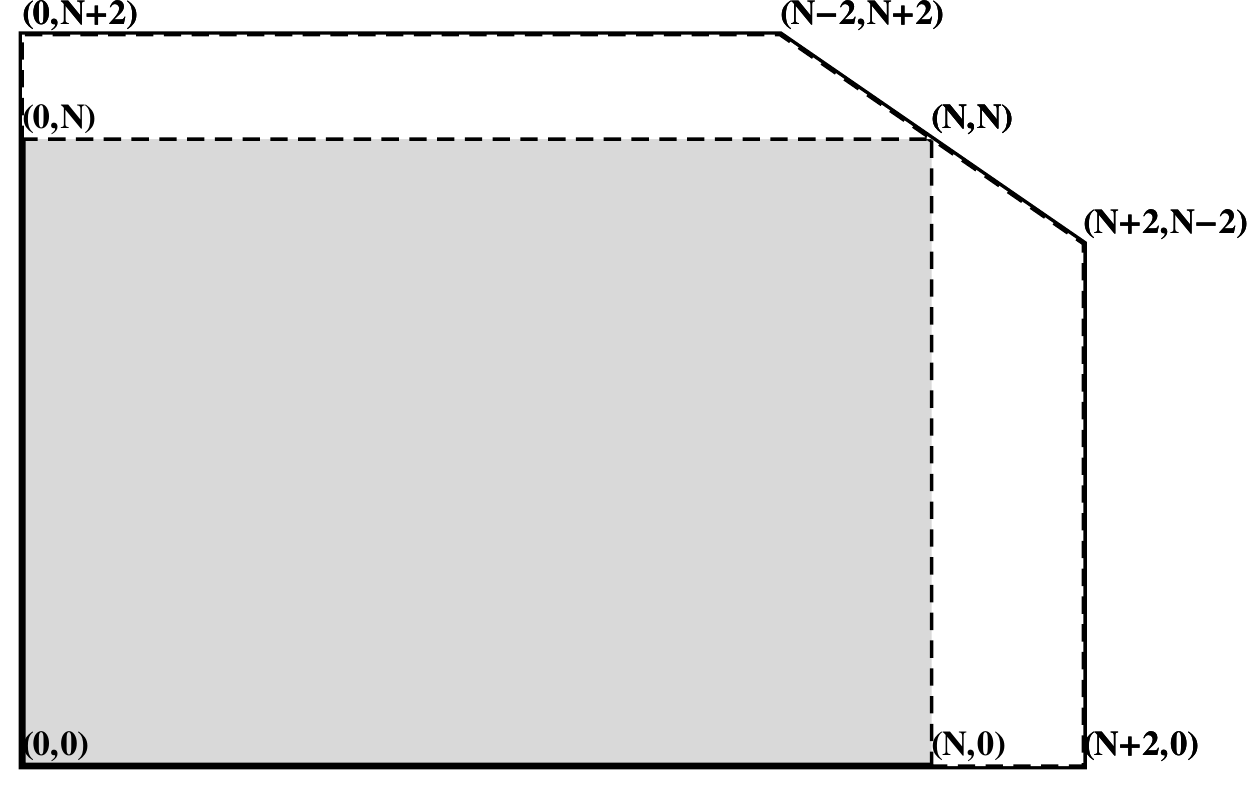}
  \caption{The Newton polygons of two polynomials $F(x,y)$ and $G(x,y)$ and their Minkowski sum is shown. The area of the the three regions are $\mbox{Area}\left(\mbox{New}\left(F(x,y).G(x,y)\right) \right) = N(N+4)-4$, $\mbox{Area}\left(\mbox{New}\left(F(x,y)\right)\right) = 2(N-1)$ and   $\mbox{Area}\left(\mbox{New}\left(G(x,y)\right)\right)= 2(N-1)$ and the mixed area  $N^2$ is given by the gray square.}

\end{figure}
In table V we give  the first type of solutions and in table VI  the second type of solutions.  By substituting  the singular roots (\ref{4sol1}) or (\ref{4sol2}) in  the Bethe ansatz  equations
(\ref{bethe}) we obtain a  polynomial equation for a single variable $x$
\begin{eqnarray}\label{bethe4root}
\left(\frac{x-i}{x+i}\right)^{N-2}= \frac{x-3i}{x+3i}\,,~~~  \mbox{for}~~~~~  N=8, 10, 12, \cdots\,,
\end{eqnarray}
where $x$ is either  $2a_1$ or  $2ia_1$, of eq. (\ref{4sol1})  and (\ref{4sol2}), respectively. This is a polynomial equation  of degree  $N-2$ which  can be seen from the simplified form 
\begin{eqnarray}\label{bethe4root1}
f(x)=A_{N-2}x^{N-2} + A_{N-4}x^{N-4} + \cdots +A_{N-2-2r}x^{N-2-2r}+ \cdots +  A_0=0\,,
\end{eqnarray}
where the coefficients  $A_{N-2}= \Comb{N-2}{1}-3, A_{N-4}=-\Comb{N-2}{3}+3\Comb{N-2}{2}, \cdots, A_0=-3(-1)^{(N-2)/2}$  are all real.   It is useful to write the general form of the coefficients  as 
\begin{eqnarray}\label{coeff}
A_{N-2-2r}= (-1)^r\left[\Comb{N-2}{2r+1}  -3\Comb{N-2}{2r}\right],~~~~ \mbox{for}~~~~~  r=0, 1, 2, \cdots, (N-2)/2\,.
\end{eqnarray}
It is manifest that if  $x$ is a root of the polynomial then $-x$ is also a root, which  accounts  for the two rapidities  of   $M=4$ singular solution in  (\ref{4sol1}) and  (\ref{4sol2}).  According to the fundamental theorem of algebra  the polynomial eq.  (\ref{bethe4root1})  has   at most  $N-2$    distinct roots (real or complex) and  since the coefficients are all real  the complex roots will occur in complex conjugate pairs if there are any.  Considering the fact that   two roots of opposite  signs  of    (\ref{bethe4root1})  constitute one singular root  we  find that  the total number of singular roots  $\mathcal{N}$ for $M=4$ is  at most
\begin{eqnarray}\label{shnroot4}
\mathcal{N}= \frac{N-2}{2}\,,~~~  \mbox{for}~~~~~  N=8, 10, 12, \cdots\,.
\end{eqnarray}
One test which guarantees  that the the total number of singular solutions is exactly  given by  (\ref{shnroot4})  is to show that the discriminant  of the polynomial  $f(x)$ in (\ref{bethe4root1})  is not zero, which we cannot prove  here.  But numerical  check for many different values of the chain lengths   shows that  discriminants are indeed  non-zero and negative, it follows that  the roots are all distinct  and all account for the singular solutions.    The  number of   sign  changes $\mathcal{V}_+$  of the coefficients  (\ref{coeff})  of  $f(x)$  and  the number of sign  changes $\mathcal{V}_-$ of the corresponding coefficient of $f(-x)$  are the  same and  given by 
\begin{eqnarray}\label{sgn12}
\mathcal{V}_{\pm}= \frac{N-2}{2}-1\,, ~~~  \mbox{for}~~~~~  N=8, 10, 12, \cdots\,.
\end{eqnarray}
Then, according to the Descartes' rule of sign, the number of real positive roots $n_+$ and and the number of real negative roots $n_-$ are bounded by 
\begin{eqnarray}\label{npm}
n_{\pm} \leq \mathcal{V}_{\pm}\,,
\end{eqnarray}
where the  upper and lower sign of  the  suffix of  left hand side  should be considered with the upper and lower sign of the  suffix of  right hand side, respectively.  Eq. (\ref{npm}) implies  that there are at most  $(N-2)/2-1$ number of solutions of the type  (\ref{4sol1}) and therefore  at least $1$  solution of the type (\ref{4sol2}). Here  and for  $M=5$  case bellow we assume that the complex root of (\ref{bethe4root1})  are all pure imaginary,  for which we do not have any analytical proof but it is supported by numerical observations up to $N=500$. 
So far our   numerical solutions shows that there is exactly  $(N-2)/2-1$  number  of solutions  of the first type  and only $1$ solution of the second type as can be seen from FIG. 1  
obtained for  $N=500$.

In a similar fashion singular solutions   for $M=5$ can be obtained, where  there exist  two different types  of singular solutions
 \begin{eqnarray}\label{5sol1}
&&\Big\{\frac{i}{2},-\frac{i}{2}, 0,a_1, -a_1\Big\}~~~~~~ \mbox{for}~~~ a_1 \in \mathbb{R}\,,\\ \label{5sol2}
&&\Big\{\frac{i}{2},-\frac{i}{2}, 0, ia_1, -ia_1\Big\}~~~~ \mbox{for}~~~ a_1\in \mathbb{R}\,.
\end{eqnarray}
In table VII we give  the first type of solutions and in table VIII and IX  the second type of solutions.  One can again substitute  (\ref{5sol1}) or (\ref{5sol2}) in the  Bethe ansatz  equations
(\ref{bethe}) to  obtain a  polynomial equation for a single variable $x= 2a_1$ or $2a_1i$
\begin{eqnarray}\label{bethe5root}
\left(\frac{x-i}{x+i}\right)^{N-2}= \frac{x-3i}{x+3i} \times \frac{x-2i}{x+2i}\,,~~~  \mbox{for}~~~~~  N= 10, 12, 14, \cdots\,.
\end{eqnarray}
Note that $x=0$ is a trivial solution of this equation which  is not a  Bethe roots. So, after factoring out  $x$ from  (\ref{bethe5root}) we again obtain a polynomial equation of the form (\ref{bethe4root1}) but now the coefficients are given by 
\begin{eqnarray}\label{coeff1}
B_{N-2-2r}= (-1)^r\left[\Comb{N-2}{2r+1} + 6\Comb{N-2}{2r-1} -5\Comb{N-2}{2r}\right],~~~~ \mbox{for}~~~~~  r=0, 1, 2, \cdots, (N-2)/2\,.
\end{eqnarray}
We can show that  the number of   sign  changes  $\mathcal{V}_+$  of the coefficients  (\ref{coeff1})  of $f(x)$  and  the number of sign  changes $\mathcal{V}_-$ of the corresponding coefficient of $f(-x)$  are the same
\begin{eqnarray}\label{sgn12}
\mathcal{V}_{\pm}= \frac{N-2}{2}-2\,, ~~~  \mbox{for}~~~~~  N=10, 12, 14, \cdots\,.
\end{eqnarray}
Applying the Descartes' rule  (\ref{npm}) now implies  that there are  at most  $(N-2)/2-2$  roots of the form  (\ref{5sol1}) and at least $2$ roots of the form (\ref{5sol2}) making the total number of singular solutions 
\begin{eqnarray}\label{shnroot5}
\mathcal{N}= \frac{N-2}{2}\,,~~~  \mbox{for}~~~~~  N=10, 12, 14, \cdots\,.
\end{eqnarray}
In FIG. 2, we see that for $N=500$, $M=5$, there are  only two solutions of the type (\ref{5sol2})  and the rest are of the form (\ref{5sol1}) and we also checked up to $N=500$ but find no exceptions.  For $M=6$, the following  four different classes  of singular solutions may exist
\begin{eqnarray}\label{sol1}
&&\Big\{\frac{i}{2},-\frac{i}{2}, a_1, -a_1, a_2, -a_2\Big\}~~~~~~~ \mbox{for}~~~ a_1, a_2 \in \mathbb{R}\,,
\\ \label{sol2}
&&\Big\{\frac{i}{2},-\frac{i}{2}, a_1, -a_1, ia_2, -ia_2\Big\}~~~~~ \mbox{for}~~~ a_1, a_2 \in \mathbb{R}\,,
\\ \label{sol3}
&&\Big\{\frac{i}{2},-\frac{i}{2}, ia_1, -ia_1, ia_2, -ia_2\Big\} ~~~ \mbox{for}~~~ a_1, a_2 \in \mathbb{R}\,,
\\ \label{sol4}
&&\Big\{\frac{i}{2},-\frac{i}{2}, a_1\pm ia_2, -a_1 \pm ia_2\Big\}~~~ \mbox{for}~~~ a_1, a_2 \in \mathbb{R}\,.
\end{eqnarray}
In table X solutions of the form (\ref{sol1}),  in Table XI  solutions of the form (\ref{sol2}),  in table  XII and table  XIV  solutions of the form  (\ref{sol3}), 
and in Table XIII  solutions of the form (\ref{sol4}) have been displayed.    A system of two variable   ($x,y$) polynomial equations for even $N\geq 12$, $M=6$ can be obtained by substituting any of the  form of the roots  (\ref{sol1})- (\ref{sol4})  in the Bethe Ansatz   equations  (\ref{bethe})  as
\begin{eqnarray}\label{bethe6aroot}
\left(\frac{x-i}{x+i}\right)^{N-2}= \frac{x-3i}{x+3i} \times \frac{x -y-2i}{x-y+2i}
\times \frac{x +y-2i}{x + y+2i}\,, \\
\left(\frac{y-i}{y+i}\right)^{N-2}= \frac{y-3i}{y+3i} \times \frac{y -x-2i}{y-x+2i}
\times \frac{y +x-2i}{y + x+2i}\,,
\end{eqnarray}
which can   be rewritten as 
\begin{eqnarray}\label{bethe6root}
\nonumber
F(x,y)=&&\sum_{r=0}^{\frac{N-2}{2}}(-1)^r\left[\left(\Comb{N-2}{2r+1}-\Comb{N-2}{2r}\right)x^{N-2r}+\left(8\Comb{N-2}{2r+1}-12\Comb{N-2}{2r}\right)x^{N-2-2r}\right.\\ \nonumber
&&\left.-\left(\Comb{N-2}{2r+1} +3\Comb{N-2}{2r}\right)x^{N-2-2r}y^2\right]=0\,,\\
\nonumber 
G(x,y)=&&\sum_{r=0}^{\frac{N-2}{2}}(-1)^r\left[\left(\Comb{N-2}{2r+1}-\Comb{N-2}{2r}\right)y^{N-2r}+\left(8\Comb{N-2}{2r+1}-12\Comb{N-2}{2r}\right)y^{N-2-2r}\right.\\ 
&&\left.-\left(\Comb{N-2}{2r+1} +3\Comb{N-2}{2r}\right)y^{N-2-2r}x^2\right]=0\,.
\end{eqnarray}
where the pair ($x,y$) can  either be ($2a_1, 2a_2$),  ($2a_1, 2ia_2$),  ($2ia_1, 2ia_2$), ($2a_1+ 2ia_2, 2a_1-2ia_2$) or ($2a_1+ 2ia_2, -2a_1+ 2ia_2$).  Note that the two equations  (\ref{bethe6root})  are symmetric with  respect to the permutations  of the variables.  Although solutions of this system of  equations give  all the  desired roots of the form 
 (\ref{sol1})- (\ref{sol4}), they also give   roots such as    $x= \pm y$ and $x \neq  \pm y^*$($\mathbb{R}(x)=\mathbb{R}(y) \neq 0$) which are not physical solutions and therefore should be discarded.  In order to calculate the number of singular solutions we have to first find out the number of solutions  of   (\ref{bethe6root}) and subtract the number of solutions of type  
 $x= \pm y$ and $x\neq  \pm y^*$($\mathbb{R}(x)=\mathbb{R}(y)\neq 0$).  According to  Bernstein's theorem \cite{bern} number of solutions of  a system of generic polynomial equations of two variables of the form $f_1(x,y)=0, f_2(x,y)=0$  in $(\mathbb{C}\backslash {0})^2$ is given by their  mixed area  $\mathcal{M}\left(\mbox{New}(f_1),\mbox{New}(f_2)\right)$, where  $\mbox{New}(f_1)$ and $\mbox{New}(f_2)$ are the Newton polygons of  $f_1(x,y)$ and $f_2(x,y)$, respectively.  By inspecting (\ref{bethe6root})  we can readily obtain the Newton polygons $\mbox{New}\left(F(x,y)\right)$ and $\mbox{New}\left(G(x,y)\right)$ of the system of equations 
 as 
\begin{eqnarray}\label{newpolygon}
\nonumber
\mbox{New}\left(F(x,y)\right): =\mbox{conv}\{(N,0), (N-2,0), \cdots, (0,0), (N-2,2), (N-4,2), \cdots, (0,2)\}\,,\\
\mbox{New}\left(G(x,y)\right) :=\mbox{conv}\{(0,N), (0,N-2), \cdots, (0,0), (2,N-2), (2,N-4), \cdots, (2,0)\}\,.
\end{eqnarray}
The  Minkowski  sum   $\mbox{New}\left(F(x,y).G(x,y)\right)$  of the two polygons can be obtained from the multiplication of the corresponding polynomials and can be written as 
 \begin{eqnarray}\label{newpolygon1}\nonumber
&&\mbox{New}\left(F(x,y).G(x,y)\right) := \\
&&\begin{matrix}
\mbox{conv} \Big \{(N,N),& (N-2,N), & \cdots, & (0,N), & (N-2,N+2),& (N-4,N+2), & \cdots, & (0,N+2),\\
  (N,N-2),& (N-2,N-2), & \cdots, & (0,N-2), & (N-2,N), & (N-4,N),& \cdots, & (0,N), \\
  \vdots  & \vdots  & \ddots & \vdots &  \vdots  & \vdots  & \ddots & \vdots  \\
  (N,0),& (N-2,0), & \cdots, & (0,0), & (N-2,2),& (N-4,2), & \cdots, & (0,2),\\
  (N+2,N-2),& (N,N-2), & \cdots, & (2,N-2), & (N,N), & (N-2,N),& \cdots, & (2,N), \\
    (N+2,N-4),& (N,N-4), & \cdots, & (2,N-4), & (N,N-2), & (N-2,N-2),& \cdots, & (2,N-2), \\  
    \vdots  & \vdots  & \ddots & \vdots &  \vdots  & \vdots  & \ddots & \vdots  \\
    (N+2,0),& (N,0), & \cdots, & (2,0), & (N,2), & (N-2,2),& \cdots, & (2,2) \Big\}\,.\\  
   \end{matrix}   
\end{eqnarray}
 Now one can  obtain the mixed area
 \begin{eqnarray}\label{mxarea} 
 \mathcal{M}\left( \mbox{New}\left(F\right).\mbox{New}\left(G\right)  \right)=  \mbox{Area}\left(\mbox{New}\left(F(x,y).G(x,y)\right) \right)-\mbox{Area}\left(\mbox{New}\left(F(x,y)\right) \right) -\mbox{Area}\left(\mbox{New}\left(G(x,y)\right) \right)= N^2\,,
 \end{eqnarray}
 which is  the area of the  gray colored square  in  FIG. 3. There are  $2(N-2)$ solutions of the form   $x= \pm y$, which can be shown analytically easily because in this case we can  reduce the Bethe ansatz  equation  to an one variable polynomial equation of degree $(N-2)$.  From the numerical observation for $N=12, 14, 16$ we see that  there are $4(N-1)$ roots which are of the form $x \neq \pm y^*$($\mathbb{R}(x)=\mathbb{R}(y) \neq 0$) and we assume this to be valid for any $N\geq 12$. For $N=16, M=6, 7$, singular solutions have been shown in Table XV, XVI  respectively.  So the  total  number of singular solutions becomes
\begin{eqnarray}\label{sing6}
 \mathcal{N}= \frac{1}{8}\left(N^2-6N+8\right)~~~~~\mbox{for}~~~~~ N=12,14, \cdots\,.
 \end{eqnarray} 
The overall factor $8$ in the denominator is the multiplicity of the singular roots.  Note that  two roots of  (\ref{bethe6root}) are considered the  same  if  upon substitution in  (\ref{sol1})- (\ref{sol4}) gives the same  singular  roots.
Similarly,  $M=7$ case can also be discussed, where there are  the following  four different class  of singular solutions
\begin{eqnarray}\label{sol12}
&&\Big\{\frac{i}{2},-\frac{i}{2}, 0,a_1, -a_1, a_2, -a_2\Big\}~~~~~~~ \mbox{for}~~~ a_1, a_2 \in \mathbb{R}\,,
\\ \label{sol22}
&&\Big\{\frac{i}{2},-\frac{i}{2}, 0,a_1, -a_1, ia_2, -ia_2\Big\}~~~~~ \mbox{for}~~~ a_1, a_2 \in \mathbb{R}\,,
\\ \label{sol32}
&&\Big\{\frac{i}{2},-\frac{i}{2}, 0, ia_1, -ia_1, ia_2, -ia_2\Big\}~~~ \mbox{for}~~~ a_1, a_2 \in \mathbb{R}\,,
\\ \label{sol42}
&&\Big\{\frac{i}{2},-\frac{i}{2}, 0, a_1\pm ia_2, -a_1 \pm ia_2\Big\}~~~ \mbox{for}~~~ a_1, a_2 \in \mathbb{R}\,.
\end{eqnarray}
A system of two variable   ($x,y$) polynomial equations  can   be obtained by substituting any of the  form of the roots  (\ref{sol12})- (\ref{sol42})  in the Bethe Ansatz   equations  (\ref{bethe})  as
\begin{eqnarray}\label{bethe7aroot}
\left(\frac{x-i}{x+i}\right)^{N-2}= \frac{x-3i}{x+3i} \times\frac{x-2i}{x+2i}\times \frac{x -y-2i}{x-y+2i}
\times \frac{x +y-2i}{x + y+2i}\,, \\
\left(\frac{y-i}{y+i}\right)^{N-2}= \frac{y-3i}{y+3i} \times\frac{y-2i}{y+2i}\times \frac{y -x-2i}{y-x+2i}
\times \frac{y +x-2i}{y + x+2i}\,,
\end{eqnarray}
where ($x,y$)  has been defined in eq. (\ref{bethe6root}).
The number  of singular solutions is given by the the formula  (\ref{sing6}) but now  $N=14, 16, \cdots$. Generally,  for  even-$N$ and even-$M$,  the singular  root  looks like
\begin{eqnarray}\nonumber
&&\Big\{\pm \frac{i}{2}, \pm a_1, \pm a_2, \cdots, \pm a_{n_1}, \pm ib_1, \pm ib_2, \cdots, \pm ib_{n_2}, \pm c_1\pm id_1,\pm c_2\pm id_2, \cdots, \pm c_{n_3}\pm id_{n_3}\Big\} \\ &&\mbox{for}~~~ a_i, b_i, c_i, d_i \in \mathbb{R}; n_1,n_2 \in[0,1,2, \cdots, \frac{M-2}{2}];n_3 \in[0,1,2, \cdots, \frac{M-2}{4}]; 2n_1+2n_2+4n_3= M-2\,,
\end{eqnarray}
and  for  even-$N$ and odd $M$, the singular  root  looks like
\begin{eqnarray}\nonumber
&&\Big\{\pm \frac{i}{2}, 0,  \pm a_1,  \pm a_2, \cdots, \pm a_{n_1}, \pm ib_1, \pm ib_2, \cdots, \pm ib_{n_2},  \pm c_1\pm id_1, \pm c_2\pm id_2, \cdots, \pm c_{n_3}\pm id_{n_3}\Big\}  \\ &&\mbox{for}~~~ a_i, b_i, c_i, d_i \in \mathbb{R}; n_1, n_2 \in[0,1,2, \cdots, \frac{M-3}{2}];  n_3 \in[0,1,2, \cdots, \frac{M-3}{4}]; 2n_1+2n_2+4n_3= M-3\,.
\end{eqnarray}
Solving this  problem for the number of singular solutions becomes  complicated as $M$ increases but a general form of the number of singular solutions can be written as 
\begin{eqnarray}\label{singM} 
 \mathcal{N}= \frac{1}{p_0}\left(N^n +  p_1N^{n-1} + p_2N^{n-2}+ \cdots +p_n \right) ~~~~~\mbox{for}~~~~\mbox{even} ~N \geq 2M\,,
 \end{eqnarray} 
where  the integer  $p_0$ is the multiplicity of the singular roots,  $p_i$'s are some integers and  for even $M$, $n=(M-2)/2$,  or for odd $M$,   $n=(M-3)/2$.  From the analysis of singular solutions up to $M=7$ we   conjecture that for even $N$ the number of  singular solutions  for even $M$ and  $M+1$   are  same.

\section{Rigged Configurations}
It has been observed  that there is a   connection between  the Bethe states and the rigged configurations \cite{kirillov,kirillov2,lule}.   It offers a nice bijection between the Bethe states and the rigged configurations   at least for a not very  long spin-$1/2$ chain. In   $N=12$  case of  the isotropic spin-$1/2$ chain, we   establish this  bijection for the   singular solutions and for  the  non self-conjugate  string solutions   comparing their   Takahashi  quantum numbers with  the riggings of  a rigged configuration. 

To understand what a rigged configuration  is and how it works let us give here a brief  account of  the basic idea behind the rigged configurations.  We keep the notations of \cite{kirillov}. This is a   Young Tableau  like object with two sets of integers, one in the left hand side of the boxes known as  vacancy numbers $P_k(\nu)$, and the other on the right hand  side of the  boxes known as riggings $J_{k,\alpha}$. Consider an eigenstate of 
the Hamiltonian  of a spin-$1/2$ chain  of length $N$ and total  $M$ down-spins in the state.  The  down-spins can be partitioned  in  different ways and each partition can be written as  $\nu= \{\nu_1, \nu_2, \cdots, \nu_s\}$ such that the parts  $\nu_i $'s  are positive integers and $s$ is the total number of parts in a particular partition.   Since  all the down-spins have been partitioned in $\nu$ it  satisfies  $\sum_{i=1}^{s}\nu_i= M$. In  the string-solution language, for example,  $M=9$ down-spins with  two  $3$-strings,  one $2$-string  and one $1$-string has a partition $\nu=\{3, 3, 2, 1\}$.  The set of vacancy numbers  $P_k(\nu)$ which need to be all non-negative in order to have a  viable configuration are  defined for a spin-$1/2$ system as follows
\begin{eqnarray}\label{rigg1}
P_k(\nu)= N- 2 \sum_{i=1}^{s} \mbox{min} \left(k,\nu_i\right)\,,
\end{eqnarray}
where $k=1, 2, \cdots$ is the length of a string under consideration.  Once a  vacancy number is obtained then one can get  a bound for the set of corresponding  riggings   $J_{k,\alpha}$ as 
\begin{eqnarray}\label{rigg2}
0\leq J_{k, 1} \leq J_{k, 2} \leq \cdots \leq J_{k, M_k}\leq P_k(\nu)\,,
\end{eqnarray}
where $M_k$ is the total number of $k$-strings in a particular   set of roots defining a state. In order to have a bijection between the  rigged configurations and the Bethe states we need to  define a flip map $\kappa$  as
\begin{eqnarray}\label{rigg3}
\kappa(J_{k, \alpha})= P_k(\nu)- J_{k, M_k-\alpha+1}\,.
\end{eqnarray}
A rigged configuration of the form $(\nu, J)$  therefore have two different classes of  configurations, one which are flip-invariant  and the other which are not flip-invariant under the transformation (\ref{rigg3}). 

Given a partition $\nu$ and a set of corresponding Bethe states it is   now  our task  to assign a  rigged configuration $\left(\nu, J\right)$  to a Bethe root.  One way to assign this  is to compare between the  riggings $J$ and the real part of the  rapidities of  Bethe roots and  assign  higher value of the real part of the  roots to higher value of the riggings as adopted in  \cite{kirillov}.  To get a mapping based on the comparison with the rapidity we have to then actually solve the  rapidities  numerically. We instead   considered  a comparison between the set of riggings of  $k$-length strings $\{J_{k, 1}, J_{k, 2}, \cdots, J_{k, M_k}\}$ in a form of  the lexicographical order  and the set of  Takahashi quantum numbers of the same $k$-length strings  $\{I^1_k,  I^2_k, \cdots, I^{M_k}_k\}$ in a form of  the inverse lexicographical order.  The  comparison starts from the  largest length strings to the  lowest length strings  in descending order.  We  assign   larger riggings to larger Takahashi quantum numbers  in the order specified in the above and obtain  a bijection  between the Bethe states and the rigged configurations.  
On the right hand side of each row  of  the Tables I to XIV we have    shown the corresponding rigged configurations for $N=12$ spin-$1/2$ chain. 

To understand the mapping,  let us consider all  solutions of the form of one $2$-string and two $1$-strings in  $N= 12, M=4$.  Since the largest string in this example is  one  $2$-string, we  first compare the Takahashi quantum numbers  of the two string  $\{I_2\} = \{2, 1, 0, -1, -2\}$
with the  riggings  of the $2$-string     $\{J_{2, 1}\} = \{4, 3, 2, 1, 0\}$ from the  left to the right. Then we compare  $28$ pairs of the  Takahashi quantum numbers for two $1$-strings,  
$\{I_1^1, I_1^2\} = \Big\{ (7/2, 5/2), (7/2, 3/2), \cdots,  (7/2, -7/2), (5/2, 3/2), (5/2, 1/2), \cdots, (5/2, -7/2), (3/2, 1/2), (3/2, -1/2), \cdots, (3/2, -7/2),$ $(1/2, -1/2), (1/2, -3/2), (1/2,-5/2), (1/2, -7/2), (-1/2, -3/2), (-1/2, -5/2), (-1/2, -7/2), (-3/2, -5/2),$ $ (-3/2, -7/2), (-5/2, -7/2)\Big\}$, with  $28$ pairs of  riggings of the two $1$-strings,  $\{J_{1,1}, J_{1,2}\}= \Big \{ (6, 6), (5, 6),$ $\cdots, (0, 6), (5, 5), (4, 5), \cdots, (0, 5),  (4, 4), (3, 4), \cdots, (0, 4), (3, 3), (2, 3), (1, 3), (0, 3), (2, 2), (1, 2), (0, 2),  (1, 1), (0, 1), (0, 0) \Big\}$  from the  left to the right.   Four such  explicit correspondences have been given  in TABLE V.
This is just one example,  the  mapping  for all  solutions  of the  $N= 12$ spin chain   will be reported  elsewhere.

\begin{table}[h]
\caption{Singular solutions for $N=16$, $M=6$}
\begin{tabular}{l@{\hskip 0.1in}r@{\hskip 0.1in}l@{\hskip 0.1in}l@{\hskip 0.1in}l@{\hskip 0.1in}}
\hline\hline
&$\pm \lambda_1$ & \hspace{0.6in}$\pm \lambda_2$ &\hspace{0.6in} $\pm \lambda_3$& \hspace{0.6in}$E$ \\
\hline
1.& $\pm$0.5i &	$\pm$2.5503138374817507i &	 $\pm$1.5000074388001383 &	-0.34011048736365523 \\
2.& $\pm$0.5i &	$\pm$1.3796268446813218 $\pm$0.5617044948437927i &	$\pm$1.3796268446813218  $\mp$0.5617044948437927i &	-1.6359517330815763 \\
3. &  $\pm$0.5i &	$\pm$1.5006551279784253i &	 $\pm$1.4468750235773047 &	-0.9272127431773511 \\
4& $\pm$0.5i &	        $\pm$1.5015054809619202i&	$\pm$0.7685335795282152  &	-1.6906915160727207 \\
5.& $\pm$0.5i&	        $\pm$1.1888973013271522	& $\pm$0.5651518180485309        &	-3.3573957459677906 \\
6.& $\pm$0.5i&	         $\pm$1.5025003507851975i&	    $\pm$0.4815333019660799   &	-2.577099709955288 \\
7.& $\pm$0.5i&	         $\pm$1.5034574565626846i&      $\pm$0.3032154126087406    &	-3.4270758848318454 \\
8.& $\pm$0.5i&  	$\pm$1.5042054296808929i&	     $\pm$0.16920613581844915  &	-4.092118420762521 \\
9.& $\pm$0.5i&   	$\pm$1.5046139317879763i&	    $\pm$0.05454496804382775	& -4.456399292620959 \\
10.& $\pm$0.5i&	$\pm$1.2188129982453162	 &  $\pm$0.3397640630514992	       &-4.3126312988210485 \\
11.& $\pm$0.5i&	$\pm$1.230393777779506	     &$\pm$0.18594643292085092	     &-5.080934412797139 \\
12.& $\pm$0.5i&	$\pm$0.7904601887000694  $\pm$0.4978723207970831i& 	$\pm$0.7904601887000694 $\mp$0.4978723207970831i &	-2.2383133111757774 \\
13.& $\pm$0.5i&	$\pm$1.2347813367503913        &	$\pm$0.059477281275134256  &	-5.50766928015777 \\
14.& $\pm$0.5i&	$\pm$0.9998878483722456i&	 $\pm$0.0020454192544636087i &	-3.666334749363233 \\
15.& $\pm$0.5i&	$\pm$0.6123545410064503	 & $\pm$0.3525626922450306 &	-5.2717066348913075 \\
16.& $\pm$0.5i&	$\pm$0.3690085276994174 $\pm$0.5000006776237689i& 	$\pm$0.3690085276994174 $\mp$0.5000006776237689i &-2.760293288634319 \\
17.& $\pm$0.5i&	$\pm$0.6182552619465121        &	$\pm$0.19131857038330927 &	-6.0708284216665955 \\
18.& $\pm$0.5i&	$\pm$0.6205808020482364	 & $\pm$0.060997149042557576 &	-6.515846827650192 \\
19.& $\pm$0.5i&	$\pm$0.3614712275510142	 & $\pm$0.19402031893529728	  &-7.103527226422651 \\
20.& $\pm$0.5i&	$\pm$0.3626356905570207	 & $\pm$0.06177909091569003	  &-7.561051834047694 \\
21.& $\pm$0.5i&	$\pm$0.1958958475087899	 & $\pm$0.062168651323811226  &	-8.406807180538566 \\
\hline
\end{tabular}
\end{table}

\begin{table}[h]
\caption{Singular solutions for $N=16$, $M=7$}
\begin{tabular}{l@{\hskip 0.02in}r@{\hskip 0.05in}l@{\hskip 0.1in}l@{\hskip 0.1in}l@{\hskip 0.02in}l@{\hskip 0.02in}}
\hline\hline
&$\lambda_0$ &$\pm \lambda_1$ & \hspace{0.6in}$\pm \lambda_2$ &\hspace{0.6in} $\pm \lambda_3$& \hspace{0.6in}$E$ \\
\hline
1.& 0.& $\pm$0.5i           &	$\pm$ 2.6298433893582316i&	 $\pm$ 1.5000914661481464i&	 -2.350055306797338\\
2.&0.& $\pm$0.5i            &	$\pm$ 2.07424802947408i  &$\pm$ 0.9999998530014925i&	-1.4199051831445788\\
3.&0. &  $\pm$0.5i          &	$\pm$ 1.4938977252774208i&$\pm$	1.0000016566330572i &	-1.162063053432737 \\
4.&0.& $\pm$0.5i            &	 $\pm$ 1.0698188196969034  $\pm$ 0.5269468241303141i &	$\pm$ 1.0698188196969034 $\mp$ 0.5269468241303141i&-3.8868975847171865\\
5.&0.& $\pm$0.5i            &	$\pm$ 1.5041698213826247i	& $\pm$ 1.221139754130827  &-3.0774346215313986\\
6.&0.& $\pm$0.5i            &	$\pm$ 1.5093657772817572i	&    $\pm$ 0.6592952083776108 &-3.9675056171559397\\
7.&0.& $\pm$0.5i            &	$\pm$ 1.515015400079057i&      $\pm$ 0.40717599380712116   & -4.91611439705291 \\
8.&0.& $\pm$0.5i            &$\pm$  1.5199350759696575i &	     $\pm$ 0.24334683943383273 &-5.748578542605349\\
9.&0.& $\pm$0.5i            &   	$\pm$ 1.5232105503989217i&	    $\pm$ 0.11488010687146354& -6.316377039659267\\
10.&0.& $\pm$0.5i&	$\pm$ 1.1253554306955575	 &  $\pm$	 0.9999913073015081i      & -2.326081561803587 \\
11.&0.& $\pm$0.5i&	$\pm$ 0.971587089884476   &$\pm$ 0.4785693257863526& -5.925091964375922\\
12.&0.& $\pm$0.5i&	$\pm$ 0.9925812662218414& 	$\pm$ 0.27359796126615227& -6.887862445685274\\
13.&0.& $\pm$0.5i&	$\pm$   1.0006714145894906    &	$\pm$ 0.12688494211216567& -7.557129929741305	\\
14.&0.& $\pm$0.5i&	$\pm$ 0.9999742359954078i &	 $\pm$ 0.594176313334094&	-3.3248247477794965\\
15.&0.& $\pm$0.5i&	$\pm$ 0.9999307341856244i	 & $\pm$ 0.35255560825562576 &	-4.338106532666103 \\
16.&0.& $\pm$0.5i&	$\pm$ 0.999771273297567i& 	$\pm$ 0.19159560709486184& -5.15371153919209\\
17.&0.& $\pm$0.5i&	$\pm$  0.9977671372001798i     &	$\pm$ 0.05965855699449792  &	-5.602542059539793 \\
18.&0.& $\pm$0.5i&	$\pm$ 0.5060655103518128 $\pm$ 0.49993998176856996i	 & $\pm$ 0.5060655103518128 $\mp$ 0.49993998176856996i& -4.592752302306033\\
19.&0.& $\pm$0.5i&	$\pm$ 0.508511716004234	 & $\pm$ 0.28203642842527493 & -8.00073400771648\\
20.&0.& $\pm$0.5i&	$\pm$ 0.5120886445583803	 & $\pm$  0.13008644102655006& -8.698635838944963\\
21.&0.& $\pm$0.5i&	$\pm$  0.28740641115499543 & $\pm$  0.131559767036299& -9.747595724152248\\
\hline
\end{tabular}
\end{table}

\section{Conclusions} 
We have observed that in the isotropic spin-$1/2$  Heisenberg model there are   some string solutions which do  not fall in the standard category of  string solutions.  These are physical solutions of the Bethe ansatz equations  where  the central rapidity  of  some of the  odd length strings  in a given Bethe state  becomes  complex contrary to the standard knowledge where  the central rapidity   of an  odd-length string is considered to be real even in the deformed strings.  Some of the   individual strings in a  Bethe state  in such scenario are no longer self-conjugate, but  collectively all the strings in the  Bethe state remain  self-conjugate.    This behavior  starts from $N=12$ case, where we see that  in  a  $M=5$ down spin sector   the central rapidity  of  one of the  two  1-strings  and the central rapidity of  a  3-string become complex conjugate to each other as shown in Table I.   In  $M=6$ down spin sector with $M_1=M_2=M_3=1$  we also observed that  the  the central rapidity  of an  1-string and the central rapidity  of a 3-string become complex conjugate, which are shown in Table II. 
To obtain these types of solutions in the string picture we  have used  the  Newton- Raphson method in Mathematica  and made use of  the roots of   Takahashi string with some modifications as the initial guess  for the  input  in the iteration method.   The number of missing solutions   in  the deformed string picture   for large $N$ is solely attributed to the  collapse of pairs of strings  in \cite{hag}  but as we observe  in our analysis that the missing string solutions  include not only the  collapsing strings but also the non self-conjugate  strings.   

Note that if one allows large deviations then the non self-conjugate strings, which  we investigated in this paper,  can always  be arranged in such a way that the individual strings become self-conjugate.  For example,  any  complex solution can be written in  terms  of  combination of  $1$-strings and $2$-strings, because complex  rapidities  always appear in complex conjugate pairs.  In  some cases, it is even possible to arrange  
the non self-conjugate  string solutions   in terms of  self-conjugate strings involving   larger than  $2$-strings.  However  we instead  opted for non self-conjugate string arrangement for the following reason:    
We constructed all the solutions  as small deformations of the  pure string solutions  so that we can use the  roots of the pure string solutions as an  initial guess  in the iteration process.  If we introduce  the pure string solutions as the initial guess, it is relatively easy to find  out the small deviations in the  iteration method as far as we have investigated.   
In the {\it string hypothesis} in the limit that the small deviations vanish,  we obtain the Takahashi quantum numbers, which have unique one-to-one correspondence with the  not collapsing  string solutions.   In our examples  of non self-conjugate string solutions  mentioned in  Tables I and II we emphasize this unique  correspondence  of the Takahashi quantum  numbers ( shown in the third column from the left ) to  the non self-conjugate string  solutions.  Just to clarify  the point, let us consider  the first solution of Table I, which is composed of  one real $1$-string ($0.18071431863183055$), one complex $1$-string ($ 0.44476350644863927 - 0.01877019940237738 i$)
 and one  $3$-string ($0.49181421369589934 + 0.9614711323790809 i, 0.44476350644863927 + 0.01877019940237738 i, 0.49181421369589934 - 0.9614711323790809 i$) with a complex center.   Once we take the  $\Delta_{\alpha a}^{j} \to 0$ limit, we recover the Takahashi quantum numbers associated with them, which are  also derived directly by making use of eq. (\ref{takahashi}).  On the other hand, if one arranges the above mentioned  solution in the form of one $1$-string  ($0.18071431863183055$), and two $2$-strings    ($0.49181421369589934 + 0.9614711323790809 i,  0.49181421369589934 - 0.9614711323790809 i$)  and ($ 0.44476350644863927 + 0.01877019940237738 i, 0.44476350644863927 - 0.01877019940237738 i$) with large deviations, then the self-conjugacy  is restored for each individual  string,  while there is  no  set of  Takahashi quantum numbers  available for this arrangement in the string picture. 
Here we recall that without Takahashi quantum numbers we do not have 
any good initial guess in the iteration method of finding the solutions to the Bethe ansatz equations, while the solution of the Takahashi equations for a given set of Takahashi quantum numbers leads to a good initial guess as far as we studied.  
The same is true if one arranges the above solution in the form of  one $1$-string  ($0.18071431863183055$) and  one  $4$-string ($0.49181421369589934 + 0.9614711323790809 i, 0.44476350644863927 + 0.01877019940237738 i, 0.44476350644863927 - 0.01877019940237738 i, 0.49181421369589934 - 0.9614711323790809 i$)  with  large deviations.   Similarly, every non self-conjugate solution  we  showed in this paper is given in the unique arrangement that is mapped to a set of the Takahashi quantum numbers. 
  
Considering the sum of the rapidities of a singular Bethe state to be zero for even-length spin chains we have  classified the singular roots and obtained a general form of the rapidities.   For $M=4, 5$ it enables us to reduce the Bethe ansatz  equations  to a polynomial  equation of one variable, which can be easily handled numerically for  large even  $N$ of the spin chain.  We showed that for $M=4$ and $M=5$ there are at most  $\mathcal{N} = (N-2)/2$ singular solutions   for  $N \geq 2M$.   For $M=6, 7$  it is possible to reduce the Bethe ansatz equations to a system of polynomial equations of two variables. Making use of  the algebraic method  we showed that the  number of singular solutions in such cases   are  at most   $\mathcal{N} = (N^2-6N+8)/8$.   For generic  values of even $N$ and even $M$ we can not  give the  number of singular solutions present but we can predict a form of the formula (\ref{singM}) for the number of singular solutions with coefficients  yet to be determined.  Our analysis on  the number of singular solutions  agrees with the value $\Comb{\frac{N-2}{2}}{\frac{M-2}{2}}$ for  even $M$ and $M+1$, which is  one of the conjecture in  \cite{kirillov}. However,  we showed with  examples that the number of singular solutions for even-$M$ and $M+1$ are the  same at least up to  $M=6$, which disagrees with   the conjecture $11$(C) of \cite{kirillov}, and there is  no forbidden   rigging  in connection with the singular solutions in the $N=12$ spin chain.  The  eigenvalues for the singular states have been obtained by making  use of  the regularization scheme  of \cite{bei,nepo,vlad1,kirillov2}. The detail derivation of the eigenvalues   and the eigenstates for  the singular solutions   has been given  in  \cite{girir}, where we made use of  the same  regularization scheme.

\section{Acknowledgement} 
The present study is partially supported by Grant-in-Aid for Scientific Research No. 24540396.
P. Giri acknowledges the financial support from JSPS.

\end{document}